\documentclass[11pt]{article}
\usepackage[margin=1in]{geometry}
\usepackage{amsmath,amsthm,amsfonts,amssymb}
\usepackage{graphicx,enumerate}
\usepackage[title]{appendix}
\usepackage[
            CJKbookmarks=true,
            bookmarksnumbered=true,
            bookmarksopen=true,
            colorlinks=true,
            citecolor=blue,
            linkcolor=blue,
            anchorcolor=red,
            urlcolor=blue
            ]{hyperref}
\usepackage{bm, verbatim,dsfont,mathtools}
\usepackage{algorithm,algorithmic}
\usepackage{color}
\usepackage{subfigure}
\usepackage{tikz}
\usepackage{todonotes}
\usepackage{paralist}
\usepackage{algorithm}
\usepackage{algorithmic}
\usepackage{bbm}
\usepackage{enumitem}

\newtheorem{theorem}{Theorem}
\newtheorem{lemma}{Lemma}

\newtheorem{corollary}{Corollary}
\theoremstyle{definition}
\newtheorem{definition}{Definition}

\newtheorem{remark}{Remark}
\newtheorem*{remark*}{Remark}

\newcommand{\argmin}{\mathop{\arg\min}}

\usepackage{xspace,prettyref}
\newcommand{\diverge}{\to\infty}

\newcommand{\expect}[1]{\mathbb{E}\left[ #1 \right]}

\newcommand{\prob}[1]{ \mathbb{P}\left\{ #1 \right\} }

\newcommand{\abs}[1]{ \left| #1 \right| }

\newcommand{\var}{\mathsf{var}}

\newcommand{\Binom}{{\rm Binom}}

\newcommand{\eg}{e.g.\xspace}
\newcommand{\ie}{i.e.\xspace}

\newrefformat{eq}{(\ref{#1})}
\newrefformat{chap}{Chapter~\ref{#1}}
\newrefformat{sec}{Section~\ref{#1}}
\newrefformat{algo}{Algorithm~\ref{#1}}
\newrefformat{fig}{Fig.~\ref{#1}}
\newrefformat{tab}{Table~\ref{#1}}
\newrefformat{rmk}{Remark~\ref{#1}}
\newrefformat{clm}{Claim~\ref{#1}}
\newrefformat{def}{Definition~\ref{#1}}
\newrefformat{cor}{Corollary~\ref{#1}}
\newrefformat{lmm}{Lemma~\ref{#1}}
\newrefformat{prop}{Proposition~\ref{#1}}
\newrefformat{app}{Appendix~\ref{#1}}
\newrefformat{hyp}{Hypothesis~\ref{#1}}
\newrefformat{thm}{Theorem~\ref{#1}}

\newcommand{\indc}[1]{{\mathbf{1}_{\left\{{#1}\right\}}}}

\newcommand{\calA}{{\mathcal{A}}}
\newcommand{\calB}{{\mathcal{B}}}
\newcommand{\calC}{{\mathcal{C}}}

\newcommand{\calE}{{\mathcal{E}}}
\newcommand{\calF}{{\mathcal{F}}}

\newcommand{\calR}{{\mathcal{R}}}
\newcommand{\calS}{{\mathcal{S}}}

\newcommand{\PGM}{{\sf PGM}\xspace}

\newcommand{\ER}{Erd\H{o}s-R\'enyi\xspace}
\renewcommand{\hat}{\widehat}

\renewcommand{\tilde}{\widetilde}

\allowdisplaybreaks[4]
\begin{document}
\title{The Power of  $D$-hops in Matching Power-Law Graphs}

\author{Liren Yu, Jiaming Xu, and Xiaojun Lin\thanks{
L.\ Yu  and X.\ Lin  are with School of Electrical and Computer Engineering, Purdue University, 
West Lafayette, USA, \texttt{yu827@purdue.edu, linx@ecn.purdue.edu}.
J.\ Xu is with The Fuqua School of Business, Duke University, Durham, USA, \texttt{jx77@duke.edu}.
L.~Yu and J.~Xu are supported by the NSF Grant IIS-1932630. 
}
}

\maketitle

\begin{abstract}
This paper studies seeded graph matching for power-law graphs. Assume that 
two edge-correlated graphs are  independently edge-sampled from a common parent graph with a power-law degree distribution.  
A set of correctly matched vertex-pairs is  chosen at random and revealed as initial seeds. Our goal is to use the seeds to recover the remaining latent vertex correspondence between the two graphs.
Departing from the existing approaches that focus on the use of high-degree seeds in $1$-hop neighborhoods, 
 we develop an efficient algorithm that exploits the
 low-degree seeds  in suitably-defined $D$-hop neighborhoods.
Specifically, we first match a set of vertex-pairs with appropriate degrees (which we refer to as the first slice)
based on the number of low-degree seeds in their $D$-hop neighborhoods. 
This significantly reduces the number of initial seeds
needed to trigger a cascading process to match the rest of graphs. 
Under the Chung-Lu random graph model with $n$ vertices,
max degree $\Theta(\sqrt{n})$, and the power-law exponent $2<\beta<3$, 
we show that as soon as $D> \frac{4-\beta}{3-\beta}$,
by optimally choosing the first slice, with high probability
our algorithm can correctly match a constant fraction of the true pairs without any error, provided with only $\Omega((\log n)^{4-\beta})$  initial seeds. Our result achieves an exponential reduction in the seed size requirement, as the best previously known result requires $n^{1/2+\epsilon}$ seeds (for any small constant $\epsilon>0$).
Performance evaluation with synthetic and real data further corroborates the improved performance of 
our algorithm. 
\end{abstract}

\section{Introduction}\label{sec:intro}
Given two edge-correlated graphs, graph matching 
aims to find a bijective mapping between their vertex sets  so that their edge sets are maximally aligned. 
It is a fundamental problem with numerous applications in a variety of fields, including  social network de-anonymization \cite{narayanan2009anonymizing}, machine learning \cite{cour2007balanced,fiori2013robust}, computer vision \cite{conte2004thirty,schellewald2005probabilistic}, pattern recognition \cite{berg2005shape,1265866}, computational biology \cite{singh2008global,kazemi2016proper} and natural language processing \cite{haghighi2005robust}.

This paper focuses on seeded graph matching, wherein an initial  set of seeds, \ie,
correctly matched vertex-pairs, is revealed as side information. This is motivated by the fact that in many real
applications, some side information on the vertex identities is available and has been successfully utilized
to match many real-world networks\footnote{For example, in social network de-anonymization, some users provide identifiable information in their service registrations or explicitly link their accounts across different social networks.}
\cite{narayanan2008robust,narayanan2009anonymizing}. 
Using seeds, we can then measure the similarity of a vertex-pair by its ``witnesses". More precisely, let $G_1$ and $G_2$ denote two graphs.  For each pair of vertices $(u,v)$ with $u$  in $G_1$ and $v$ in $G_2$, a seed $(w,w')$ is called a \emph{1-hop witness} for $(u,v)$ if $w$ is a neighbor of $u$ in $G_1$ and $w'$ is a neighbor of $v$ in $G_2$. Since $G_1$ and $G_2$ are  graphs with correlated edges, 
a  candidate  pair  of  vertices  are expected to have more witnesses if they are a true pair than if they are a fake pair. This idea has been applied to many graph matching problems, and strong performance guarantees (in term of the required number of seeds) have been obtained, in particular, for matching \ER graphs~\cite{pedarsani2011privacy,yartseva2013performance,Lyzinski2013Seeded,korula2014efficient,10.14778/2794367.2794371,shirani2017seeded,Fishkind2018Seeded,mossel2019seeded,lubars2018correcting,yu2021graph}.


However, 
\ER graphs fall short of capturing many fundamental structural properties of
real-world networks. Notably, many real-world networks 
exhibit a power-law degree distribution, \ie,
the fraction of nodes with degree at least $k$ decays as $k^{-\beta+1}$ for some exponent $\beta>0.$
As a consequence, we expect to see  very large degree fluctuations, with 
some nodes having very high degrees (so-called
hubs) and some other sparsely-connected  nodes  with small degrees. Intuitively, this degree fluctuation may confuse witness-based vertex matching, \eg, a fake pair with high degrees may have many more witnesses than a true pair with low degrees, 
which foils 
the existing seeded algorithms designed for matching \ER graphs.

There have been several attempts to design seeded graph matching algorithms for power-law graphs
\cite{korula2014efficient,10.1109/TNET.2016.2553843, 10.1007/s00453-017-0395-0}. 
However, they tend to require a larger number of seeds than \ER graphs. Note that to address the above-mentioned degree variations, a common idea is to first partition graphs into slices  consisting of
vertex-pairs with similar degrees.
Then, the vertices are matched slice-by-slice, starting from the highest-degree slice to lower-degree slices. A cascade process is triggered,
in the sense
that the matched vertices in the current slice is used as new seeds to match the next slice. 
Intuitively, it is critical to correctly match the first slice in order to successfully trigger 
the cascading matching process for the later slices. \cite{korula2014efficient,10.1109/TNET.2016.2553843, 10.1007/s00453-017-0395-0} all use this idea and match the first slice based on 1-hop witnesses. Unfortunately, they also require a large number of correct seeds to match the first slice successfully. 
Specifically, \cite{korula2014efficient} assumes 
preferential-attachment graphs with $n$ vertices \cite{barabasi1999emergence}  and their algorithm requires  $\Omega(n/\log(n))$ seeds to match a constant fraction of all vertices correctly. \cite{10.1109/TNET.2016.2553843, 10.1007/s00453-017-0395-0} instead assume the Chung-Lu graph model \cite{chung2006complex} (cf.~\prettyref{sec:model}). 
When all seeds are chosen from the high-degree vertices, \cite{10.1109/TNET.2016.2553843, 10.1007/s00453-017-0395-0} show that their algorithm require only $n^\epsilon$ seeds to correctly match a constant fraction of the vertices. However, if the seeds are chosen uniformly from all vertices, the number of high-degree seeds will be much smaller than $n^\epsilon$. In that case, the degree-driven graph matching (DDM) algorithm in \cite{10.1109/TNET.2016.2553843} requires $n^{1/2+\epsilon}$ seeds to match a constant fraction of vertices correctly.

In this paper, we propose a new algorithm for matching power-law graphs that only requires $\Omega((\log n)^{4-\beta})$ initial seeds chosen randomly, to correctly match a provably constant fraction of all vertices. Our key departure from \cite{korula2014efficient,10.1109/TNET.2016.2553843, 10.1007/s00453-017-0395-0} is to use ``witnesses" in larger $D$-hop neighborhoods. 
More precisely, a seed $(w,w')$ is a $D$-hop witness for $(u,v)$ if $w$ is a $D$-hop neighbor of 
$u$ in $G_1$ and $w'$ is a $D$-hop neighbor of $v$ in $G_2. $
To see why using $D$-hop witnesses is crucial, note that, under the Chung-Lu model of  \cite{chung2006complex} (cf. \prettyref{sec:model}), even the highest degree vertices only have a 1-hop neighborhood of size at most $O(\sqrt{n})$. Since seeds are uniformly chosen, it is clear that at least $\Omega(\sqrt{n})$ seeds are needed to ensure that a true pair in the first slice can have $\Omega(1)$ 1-hop witnesses. In contrast, 
as $D$ increases, the size of the $D$-hop neighborhoods grows rapidly,
and thus there are substantially more seeds that can serve as $D$-hop witnesses for true pairs, which provides hope to significantly reduce the number of initial seeds. 

The idea of $D$-hop witnesses has also been used for matching \ER graphs in \cite{mossel2019seeded,yu2021graph}. However, as can be seen in the rest of the paper, the application of $D$-hop witnesses to power-law graphs is highly non-trivial.  Specifically, due to the power-law degree variations, the $D$-hop neighborhoods of some high-degree
vertices may become so large that even a fake pair can have many $D$-hop witnesses. 
Therefore, a key challenge is to properly control the size of the $D$-hop neighborhoods. This size depends not only on the degrees of the vertex pair to be matched, but also that of the intermediate nodes (to reach $D$-hop) and that of the seeds. 
To overcome this challenge, our algorithm design (to be explained in \prettyref{sec:idea}) (i)  carefully chooses the first slice of vertices to be matched. (ii) carefully chooses the intermediate vertices when
constructing the $D$-hop neighborhoods, and (iii) carefully avoids high-degree seeds in order to eliminate the confusion for fake pairs.
These three ideas altogether ensure that the true pairs in the first slice have many more $D$-hop witnesses 
than the fake pairs, and thus can be correctly matched to trigger the cascading process
to match the rest of the graphs. See~\prettyref{sec:idea} for more detailed discussions. 

To fully realize the power of $D$-hops, we further need to carefully construct overlapping 
slices to account for the potential mismatch in the vertex slicing of
graphs $G_1$ and $G_2$, and to design effective ways to 
match the remaining slices. 
Assembling all these pieces together enables  us 
to achieve an exponential reduction in the required number of seeds compared to state-of-art results in \cite{10.1109/TNET.2016.2553843}. Specifically, 
under the Chung-Lu model with power-law exponent $2<\beta<3$ and max degree $\Theta(\sqrt{n})$, we prove the following
performance guarantee of our algorithm,  stated informally here and formally in \prettyref{sec:result}. 
\begin{theorem}[Summary of main result]\label{thm:summary}
Suppose
$D > \frac{4-\beta}{3-\beta}$.  
If there are $\Omega((\log n)^{4-\beta})$ 
initial seeds chosen independently at random, by optimally choosing the first slice, 
our algorithm correctly matches $\Omega(n)$ vertex-pairs without any error with high probability.
\end{theorem}
 This reduces the seed size requirement exponentially, as the best previously known result~\cite{10.1109/TNET.2016.2553843} requires $n^{1/2+\epsilon}$ seeds. 
To prove~\prettyref{thm:summary}, there are several key innovations
in our analysis in particular to address the difficult dependency issues across edges and slices. First, note that when we define the $D$-hop neighborhoods,  we use vertex degrees
to construct the slices and 
to select 
the seeds and intermediate nodes.
This degree-based slicing unfortunately 
brings dependency issues. In particular, if we condition on the vertex degrees,
then the edges are no longer independently generated according to the Chung-Lu model.
To circumvent this dependency issue, we first show that the 
degree-guided construction
and selection can be closely approximated by the weight-guided counterparts
with high probability. Then we restore the independence
by studying the weight-guided
construction and selection, since the edges are independently generated 
according to the Chung-Lu model given the weights. 
Second, as we use the matched pairs in the current slice as new seeds
to match the next slice, the matching results are correlated across different slices.
To deal with these correlations, we carefully construct 
sets of matched pairs that only depend on vertex weights
to ``sandwich'' the original set of matched pairs at each slice, but are not correlated any more, which allows us to eliminate the slice-dependency issue. 
Last but not least, to derive the optimal choice of the first slice and 
attain the smallest seed size requirement,
we tightly bound the sizes 
of the common $D$-hop neighborhoods for both true pairs and fake pairs.
 Compared
to the \ER graphs, this requires much more sophisticated lines of
analysis of the neighborhood exploration
process in the power-law graphs due to the heterogeneous vertex weights.


In the literature, the idea of $D$-hop witnesses has been used in \ER graphs \cite{mossel2019seeded,yu2021graph}. However, there is a significant difference in our results for  power-law graphs. Specifically, 
in the \ER graphs with average degree $d$, 
the sizes of the $D$-hop neighborhoods are highly concentrated on $d^D$. Moreover, when the average degree $d$ is
a constant, the size of $D$-hop neighborhoods is always $O(1)$ for any constant $D$. Thus, unless $D$ increases with $n$, at least $\Omega(n)$ seeds are still needed to ensure that there are enough $D$-hop witnesses for true pairs. In stark contrast,  the power of the $D$-hop becomes much more significant for matching power-law graphs. In particular,
for power-law graphs with constant average degrees, by properly using the $D$-hop witnesses, 
we dramatically reduce the seed requirement to $\Omega((\log n)^{4-\beta})$, 
as soon as $D$ exceeds $\frac{4-\beta}{3-\beta}$. Further, we note that the algorithms in \cite{mossel2019seeded,yu2021graph} do not need to worry about controlling the $D$-hop neighborhood, as they do not face the challenge of power-law degree variations.

Finally, we conduct extensive experiments on both synthetic and real power-law graphs to corroborate our theoretical analysis. In particular, we compare our algorithm
with five other state-of-the-art seeded graph matching algorithms. 
Numerical results demonstrate that our algorithm drastically boosts the matching accuracy and
requires substantially  fewer seeds to correctly match a large fraction of vertices. 
Further, although our analysis focuses on matching two graphs of the same number of vertices, our algorithm can be  readily applied to 
match two graphs of different sizes and return an accurate matching between vertices in the common subgraph of the two graphs. Indeed, our experiments on real networks in \prettyref{sec:exp-facebook} and~\prettyref{sec:exp-auto} show that our algorithm still achieves outstanding matching performance, even when two graphs are of very different sizes.

\section{Model}\label{sec:model} 

Following~\cite{10.1109/TNET.2016.2553843, 10.1007/s00453-017-0395-0}, we adopt the
Chung-Lu random graph model \cite{chung2006complex} to generate the underlying 
parent graph with a power-law degree distribution.
Here, $[n]$ denotes the set $\{1,2,...,n\}$.

\begin{definition}
Given parameters $\overline{w}>0,$ $ \overline{w} \ll w_{\max} \le \sqrt{n\overline{w}},$ 
and  $\beta>2$, 
the Chung-Lu graph is a random graph $G_0([n],E)$ generated as follows. 
Each vertex $i\in [n]$ is associated with a positive weight $w_{i}=\overline{w}\frac{\beta-2}{\beta-1}\left(\frac{n}{i+i_0}\right)^{\frac{1}{\beta-1}}$,
where $i_0=n\left(\frac{\overline{w}(\beta-2)}{w_{\max}(\beta-1)}\right)^{\beta-1}$.
For any pair of two vertices $i,j \in [n]$ with $i\neq j$, they are connected independently by
an edge  with probability $p_{ij} =\frac{w_iw_j}{n\overline{w}}$.
\end{definition}

Note that $i_0$ is chosen such that $w_0=w_{\max}$, which is the largest weight among all vertices. Further, $\overline{w}$ approximates the average weight as follows.  Since $\overline{w} \ll w_{\max}$,
it follows that $i_0 \ll n$.
It can then be verified that $\frac{1}{n} \sum_{i=1}^n w_i \to \overline{w} $
and $\frac{1}{n} \sum_{i=1}^n \indc{w_i \ge w} \propto w^{-\beta+1}$
as $n \to \infty.$\footnote{To see the first part of the statement, let $f(x)=w_x/n$.  
Then $\int_1^{n+1} f(x)dx \le \frac{1}{n} \sum_{i=1}^n w_i \le f(n)+ \int_1^n f(x)dx$. 
Moreover,
$
 \int_1^n f(x) d x
 =\overline{w}n^{\frac{2-\beta}{\beta-1}} \bigg( (n+i_0+1)^{\frac{\beta-2}{\beta-1}} -  (i_0+1)^{\frac{\beta-2}{\beta-1}}\bigg) \to w,
 $
 in view of $i_0\ll n$ due to $w_{\max}\gg \overline{w}$.
 Further, we can verify the second part of the statement by 
 $\frac{1}{n} \sum_{i=1}^n \indc{w_i \ge w} = 
 \left( \frac{(\beta-2) \overline{w} }{(\beta-1) w} \right)^{\beta-1} - \frac{i_0}{n}$
 $\to \left( \frac{(\beta-2) \overline{w} }{(\beta-1) w} \right)^{\beta-1}.$
 } 
Thus, the degree of vertex $i$ is expected to be close to $w_i$,
which admits a power-law distribution with exponent $\beta.$

The Chung-Lu model is convenient for modelling the degree variations in real-world networks. In these real-world networks, while the average degree is often a constant,
a small but non-negligible fraction of the vertices has very large degrees (the so-called hubs) \cite{barabasi2016network}.
To model such sparse power-law graphs with hubs, we assume  $\overline{w}=\Theta(1)$ and $2 <\beta< 3$. Empirical studies have shown that the vertex degrees of many  real-world networks indeed follow a power-law distribution with $2 <\beta <3$~\cite{barabasi2016network,clauset2009power,newman2003structure}.
Note that if $0<\beta\leq 2$,  the average degree diverges and the network cannot be
sparse; if $\beta\geq 3$,  the degree variance is bounded and no large hub can appear \cite{barabasi2016network}. 

Next, we obtain a subgraph $G_1$ by sampling each edge of $G_0$ into $G_1$ independently with probability $s$, which is a constant independent of $n$. To construct another subgraph $G_2$, repeat the same sub-sampling process independently and relabel the vertices according to an \emph{unknown} permutation $\pi:[n]\to [n]$. 
Throughout the paper, we denote a vertex-pair by $(u,v)$, where $u\in G_1$ and $v\in G_2$. For each vertex-pair $(u,v)$, if $v=\pi(u)$, then $(u,v)$ is a true pair; if $v\neq \pi(u)$, then $(u,v)$ is a fake pair.

Finally, there is an initial seed set $\calS$ consisting of true pairs. Each true pair is added into $\calS$ with probability $\theta$ independently. Our goal is to recover $\pi$ based on the observation of $G_1, G_2$ and $\calS$. 

\paragraph*{Notation} 
We use standard asymptotic notation: for two positive sequences $\{a_n\}$ and $\{b_n\}$, we write $a_n = O(b_n)$ or $a_n \lesssim b_n$, if $a_n \le C b_n$ for some an absolute constant $C$ and for all $n$; $a_n = \Omega(b_n)$ or $a_n \gtrsim b_n$, if $b_n = O(a_n)$; $a_n = \Theta(b_n)$ or $a_n \asymp b_n$, if $a_n = O(b_n)$ and $a_n = \Omega(b_n)$; $a_n = o(b_n)$ or $b_n = \omega(a_n)$, if $a_n / b_n \to 0$ as $n\diverge$.

\section{Key Algorithmic Ideas}\label{sec:idea}

As we discussed in \prettyref{sec:intro},  previous graph matching algorithms that use 1-hop witnesses to match power-law graphs \cite{korula2014efficient,10.1109/TNET.2016.2553843, 10.1007/s00453-017-0395-0} require at least $n^{1/2+\epsilon}$ seeds if the seeds are chosen uniformly from all vertices. In order to significantly reduce the number of seeds, it is then crucial to use $D$-hop witnesses. However, for power-law graphs, the use of $D$-hop witnesses is highly non-trivial because, as  the size of the $D$-hop neighborhood increases, fake pairs may also have many $D$-hop witnesses and thus could be confused as true pairs. Therefore, it is important to carefully control the $D$-hop neighborhood. In this section, we elaborate on our three design choices to 
properly control the $D$-hop neighborhood sizes:
the weight of the seeds, the weight of the candidate vertex-pairs, and the weight of the intermediate vertices.


First, it is important to utilize low-weight seeds while avoiding high-weight seeds. Due to the power-law degree distribution,
when seeds are uniformly chosen,  
there are many more low-weight seeds than high-weight seeds. Thus, the $D$-hop neighborhoods need to be large enough to reach sufficiently many low-weight seeds. However, for fake pairs, their large $D$-hop neighborhoods may also overlap. This implies that high-weight seeds may easily become witnesses for fake pairs, which can appear in many $D$-hop neighborhoods. Therefore, in order to avoid having too many witnesses for fake pairs, it is important to eliminate the high-weight seeds. 
 
Second, for a given $D$, we need to carefully choose the first slice of candidate vertex-pairs
to be matched using the $D$-hop witnesses. Recall that \cite{korula2014efficient,10.1109/TNET.2016.2553843, 10.1007/s00453-017-0395-0} also use this idea of slicing the vertices according to their degrees, and focus on matching the first slice with vertices of high degree. However, we find that the degree range of this first slice needs to be carefully chosen.  
On the one hand, if the weight of the candidate vertex-pairs is too small, the common $D$-hop neighborhoods of a true pair are too small to produce enough witnesses. 
On the other hand, if the weight of the candidate vertex-pairs is too large, the $D$-hop neighborhoods of a fake pair would intersect a lot, leading to too many $D$-hop witnesses. 



Third, the high-weight vertices are not suitable to be the intermediate vertices in $D$-hop neighborhoods when $D$ is large. This is because, when $D$ is large, there exist some high-weight vertices with very large $d$-hop ($d<D$) neighborhoods. If these high-weight vertices become $(D-d)$-hop neighbors of the candidate vertices, the $D$-hop neighborhoods of the fake pairs would become too large. Thus, we should avoid using the high-weight vertices as the intermediate vertices.

Prompted by the above three ideas, we partition the graph into ``perfect" slices 
\begin{align}\label{eq:perfectslice}
  {P}_k=\{ u:w_u\in[ \alpha_{k},\alpha_{k-1}]\}  \quad \text{ where } \alpha_k=n^{\gamma}/2^k
\text{ for } k \ge 0,\text{ and } \alpha_{-1}=\infty,   
\end{align}
for some $\gamma\in(0,\log_n w_{\max}]$. In particular, the first slice $P_1$ is the set of vertices with weight in $[n^{\gamma}/2, n^{\gamma}]$, which is the first set of the vertices that we wish to match. We will show in~\prettyref{eq:intuitionof-dhoptrue}
that for a vertex in the first slice $P_1$, 
its number of $\Theta(1)$-weight $D$-hop  neighbors is on the order of $n^{\gamma((3-\beta)(D-1)+1}$.
Hence, we optimally choose $\gamma$ close to $\frac{1}{ (3-\beta)(D-1)+1}$ so that 
its number of $\Theta(1)$-weight $D$-hop  neighbors is close to $\Theta(n).$
Under this optimal choice, we prove that sufficiently many vertex-pairs in the first slice are correctly matched 
so that they can be used as new seeds to trigger the cascading process to match the rest of the graphs slice-by-slice. In fact, for slice $k\ge 2$ until $k=k^*$
for some $k^*$,  since the earlier slices provide so many new seeds, it turns out that using $1$-hop witnesses suffices. 
When $k> k^*$, the slice-by-slice matching process stops, as there are not enough $1$-hop witnesses
to correctly match the slices with low-weight vertices. Fortunately,  for the fake pairs with such low-weights,
there are very few $1$-hop witnesses as well. 
Thus we treat all the low-weight vertices as a single slice and 
apply the PGM algorithm in \cite{10.14778/2794367.2794371} to match them. 
Finally, we use all the matched vertex-pairs as new seeds 
to match the zero slice $P_0$ with very high weights. 

For the above ideas to work, however, it is important that the earlier slices do not produce wrong matches; otherwise, the wrong matches will propagate errors to the subsequent slices. As such, we only match pairs with the number of witnesses larger than a threshold, as we will see next in the detailed algorithm.

\section{The Power-Law $D$-hop (PLD) Algorithm}

In this section, we present our Power-Law D-hop (PLD)  algorithm, shown in Algorithm \prettyref{alg:D-hop-power-law} and provide the intuition why it works.

\subsection{Algorithm description}
We first introduce some notations regarding $D$-hop neighborhoods.
Given any graph $G$ and two vertices $u,v$ in $G$, we denote the length of the shortest path from $u$ to $v$ in $G$ by $\text{dist}_G(u,v)$. For each vertex $u\in G$, the $d$-hop neighbors of $u$ is denoted by $\Gamma_{d}^{G} (u)=\left\{v\in G: \text{dist}_G(u,v)=d \right\}$. The neighbors within $d$-hop of $u$ is denoted by $N_{d}^{G} (u)=\bigcup_{j=1}^{d}\Gamma_{j}^{G} (u)$.


Our PLD algorithm carefully incorporates the key algorithmic ideas described in Section~\ref{sec:idea}. At a high-level, we first slice the vertices according to their degrees. We then apply the $D$-hop algorithm to the first slice (which is carefully chosen). Afterwards, we apply the $1$-hop algorithm to the lower-degree slices $2$ to $k^*$, until the vertex degrees are about poly-logarithmic in $n$, in which case we apply the PGM algorithm to the last slice with the lowest-degree vertices. Finally, we return to slice 0 of vertices with very high degrees.  

The full algorithm is presented in Algorithm \prettyref{alg:D-hop-power-law}. We now describe the details. 

\begin{algorithm}[h]
 \caption{The Power-Law D-hop (PLD) Algorithm.}
  \begin{algorithmic}[1]
  \STATE \textbf{Input:} Graphs $G_1$ and $G_2,$ initial seed set $ \calS$, parameters $D,\gamma,\tau_1,\tau_2, k^*$
  \STATE Construct a subset of low-degree seeds $\hat{\calS}=\left\{(u,v)\in \calS: \abs{\Gamma_1^{G_1}(u)},\abs{\Gamma_1^{G_2}(v)}\leq 5\log n\right\}$.
  \STATE  Let $\hat{G}_i$ denote the subgraph of $G_i$ induced by the vertex set $V_i=\left\{u: \abs{\Gamma_1^{G_i}(u)}\leq (1+\delta) n^{\gamma}s\right\}$ for $i=1,2.$
    \STATE Partition the graph $G_i$ into slices $\hat{P}_k^{G_i}$ for $i=1,2$ and $0\le k\le k^*$, according to \prettyref{eq:imperfect_slice}.
  \STATE In $\hat{G}_1$ and $\hat{G}_2$, for  candidate vertex-pairs in $\hat{Q}_1$, count their $D$-hop witnesses in $\hat{\calS}$ and use GMWM to match
  pairs with more than $\tau_1$ $D$-hop witnesses ($\tau_1$ is given in \prettyref{eq:def_tau_1}). The set of matched pairs is $\calR_1$.
  \FOR {$k=2$ to $k^*$}
  \STATE For candidate vertex-pairs in $\hat{Q}_k$, count their $1$-hop witnesses in $\calR_{k-1}$ and use GMWM to match pairs with more than $\tau_2(k)$ $1$-hop witnesses ($\tau_2(k)$ is given in \prettyref{eq:def_tau_2}). 
  The set of matched pairs is $\calR_k$.
  \ENDFOR
  \STATE Let $G'_i$ denote the subgraph of $G_i$ induced by the vertex set 
  $V'_i=\left\{u: \abs{\Gamma_1^{G_i}(u)}\leq (1+\delta) \alpha_{k^*-1}s\right\}$, for $i=1,2.$
  \STATE Apply \PGM to $G'_1$ and $G_2'$, with the seed set $\calR_{k^*}$ and the threshold $r=3$. The set of matched pairs is denoted by $\calR_{k^*+1}$.
  \STATE For candidate vertex-pairs in $\hat{Q}_0$, count their $1$-hop witnesses in $\hat{\calR}\triangleq \bigcup_{k=1}^{k^*+1}\calR_k$  and match pairs with GMWM. The set of matched pairs is $\calR_{0}$.
  \STATE \textbf{Output:} All matched pairs $\calR=\hat{\calR}\cup\calR_0\cup \calS$
  \end{algorithmic}\label{alg:D-hop-power-law}
\end{algorithm}

In line 2, we construct a subset of low-weight seeds to use a future witnesses. Recall from \prettyref{sec:idea} that we aim to utilize low-weight seeds while avoiding high-weight seeds. Specifically, in our algorithm we wish to use seeds with $\Theta(1)$ weights. However, since we do not have access to the vertex weights directly, we have to estimate vertex weights by vertex degrees. Therefore, we construct a seed subset $\hat{\calS}$ that contains seeds with degrees no larger than $5\log n$ to ensure that all seeds with $\Theta(1)$ weights are included. 

In line 3, we eliminate the vertices with degrees larger than $(1+\delta)n^{\gamma}$ and their adjacent edges, because we do not want to use the high-weight vertices as the intermediate vertices. 

In  line 4, we partition the graphs $G_1$ and $G_2$ into slices. Recall that the ``perfect'' slices $P_k$ in \prettyref{eq:perfectslice} described in Section~\ref{sec:idea} are defined with the vertex weights. Again, since we can not observe the vertex weight directly, we need to use the vertex degree as an estimate of the vertex weight. However, using vertex degree to slice vertices creates new technical difficulties. 
Specifically, for two vertices corresponding to a true pair, their actual degrees in $G_1$ and $G_2$ may differ because the edges are sub-sampled from the parent graph randomly. As a result, it is possible that these two vertices are assigned to two slices of different indices in $G_1$ and $G_2.$ This becomes problematic because, if we only match slices with the same index, such a true pair would never be matched. (This problem does not occur for the ``perfect'' slices since they are based on the weight of the vertex in the original parent graph.)
Fortunately, the actual degrees of the vertices corresponding to a true pair should not differ too much (assuming a common sub-sampling probability $s$ for both graphs). Thus, to address the above difficulty, we enlarge the slices a little bit, so that with high probability the two vertices corresponding to a true pair can fall into slices with the same index, and therefore have the opportunity to be matched.  
More precisely, for $k\ge 0$, we define the imperfect slice as \begin{align}
    \hat{P}_k^G=\left\{u: (1-\delta) \alpha_{k}s\leq\abs{\Gamma_1^{G}(u)}\leq(1+\delta)\alpha_{k-1}s\right\}, \text{ for } k\ge 0, \label{eq:imperfect_slice}
\end{align}
where $\delta=\frac{1}{8}$ throughout this paper.
Here, $\alpha_k$ are the same as \prettyref{eq:perfectslice}, and the parameters $\gamma$ and $D$ will be set to satisfy \prettyref{eq:locallytreelikecondition} in \prettyref{thm:ChungLutheta}. 
The imperfect slice-pair is then defined as $\hat{Q}_k=\hat{P}_k^{G_1}\times\hat{P}_k^{G_2}=\{(u,v):u\in \hat{P}_k^{G_1}, v\in \hat{P}_k^{G_2}\}$. However, these enlarged imperfect slices also create a new problem of matching fake pairs, which will be discussed next. 

In line 5, we count the $D$-hop witnesses for all vertex-pairs in the first slices $\hat{P}_1^{G_1}$ and $\hat{P}_1^{G_2}$,
and then use Greedy Maximum Weight Matching (GMWM) \cite{avis1983survey} to find the vertex correspondence  such that the total number of witnesses is large. Here, we note that our earlier idea of enlarging the imperfect slices $\hat{P}_k$ creates a new problem. That is, the imperfect slices with neighboring indices now have some overlap. As a result, it is possible that a slice pair contains a fake pair $(u,\pi(v))$, but does not contain the true pairs $(u,\pi(u))$ and $(v,\pi(v))$\footnote{This phenomenon does not contradict the idea of enlarging the slices. Enlarging the slices  only guarantees the true pairs $(u,\pi(u))$ and $(v,\pi(v))$ are assigned into some slice-pairs. However, for other slice-pairs that contain the fake pair $(u,\pi(v))$, it is still possible that the two true pairs are not included.}. When that happens, the fake pairs $(u,\pi(v))$ may have the most witnesses among all the candidate vertex-pairs containing either $u$ or $\pi(v)$. Thus, the fake pair $(u,\pi(v))$ may be matched by GMWM. Fortunately, the number of witnesses of these fake pairs is still expected to be smaller than that of any true pair. Therefore, to resolve this difficulty and to ensure that only the true pairs are matched, for the first slice we match only the vertex-pairs with no less than $\tau_1$ $D$-hop witnesses, where $\tau_1$ is set to be a constant fraction of the expected number of the $D$-hop witnesses for true pairs, \ie, 
\begin{align}
    \tau_1=\frac{3}{10}\left(\frac{Cs^2}{12\overline{w}}\right)^Dn^{\gamma((3-\beta)(D-1)+1)}\theta, \label{eq:def_tau_1}
\end{align}
 where $C\triangleq(2^{\beta-1}-1)\left(\frac{(\beta-2)\overline{w}}{(\beta-1)}\right)^{\beta-1}$. 
Similar thresholds are also used in the following steps when we match other slices. 


In line 6-8,  we use the matched pairs from the previous slice as new seeds, and use the 1-hop algorithm to match  the vertices in slices $k=2,...,k^*$, where 
\begin{align}
k^*=\left\lfloor\log_2\left(n^{\gamma}\left(\frac{Cs^2}{192\overline{w}\log n}\right)^{\frac{1}{3-\beta}}\right)\right\rfloor. \label{eq:choice_k_star}
\end{align}
In other words, we match the vertices with degrees larger than $(1-\delta) \alpha_{k^*}$, where $\alpha_{k^*}\geq\left(\frac{192\overline{w}\log n}{Cs^2}\right)^{\frac{1}{3-\beta}}$. 
Again to ensure that only the true pairs are matched for each slice, we only match the vertex-pairs with at least
$\tau_2(k)$ $1$-hop witnesses, where $\tau_2(k)$ is set to be half of 
the expected number of the 1-hop witnesses of the true pairs, \ie,
\begin{align}
\tau_2(k)=\frac{C\alpha_{k-1}^{3-\beta}s^2}{16\overline{w}}.\label{eq:def_tau_2}
\end{align}

In line 9-10, 
we apply the PGM algorithm \cite[Section~3]{yartseva2013performance} to match the remaining vertices with degrees no larger than $(1+\delta)\alpha_{k^*}$. Note that when the vertex weight is this small, estimating the vertex weight based on its degree is not accurate anymore. Thus, it is difficult to use the vertex degree to distinguish which slices should these vertices fall into.
Instead, we treat all of these low-weight vertices as one slice. 
Further, for such low-degree vertices, using 1-hop algorithm based on the seeds from earlier slices will lead to poor performance, because even the true pairs in this slice have too few 1-hop witnesses. 
Fortunately, there are even fewer witnesses for the fake pairs with such low degrees. Thus, we can use the PGM algorithm, which iteratively generates new seeds as new correct matches are found. In this way, PGM can match a constant fraction of the rest of vertex-pairs, while avoiding matching fake pairs.

Finally, in line 11, the algorithm uses all vertex-pairs matched above as new seeds and matches the vertices in $\hat{Q}_0$ via the 1-hop algorithm. 

The total complexity of our algorithm is $O(n^{3-2\gamma(\beta-1)})$.  The  proof can be found in Appendix \ref{sec:complexity}.



\subsection{Intuition}\label{sec:intuition}
Before we present the main results, we explain the intuition why the above algorithm will work only with $\Omega((\log n)^{4-\beta})$ seeds. For the purpose of explaining this
intuition, we ignore the inaccuracy of estimating the weights by the vertex degrees
and assume that the graphs can be partitioned into perfect slices $P_k$. 
We further assume  that the true mapping  $\pi$ is the identity permutation. 
Also, when we write $\approx$, we ignore the constant factors that are non-essential. 



The key to the success of Algorithm \ref{alg:D-hop-power-law} is appropriately choosing
the first slice to apply the $D$-hop algorithm. 
We first calculate the probability that a vertex of $\Theta(1)$ weight lies in the $D$-hop neighborhood of a vertex in the first slice.
Specifically, given a vertex $u$ in the first slice ${P}_1$ and another vertex $v$ of weight $1$, we want to compute the probability $q_D$ that $v$ is a $D$-hop neighbor of $u$, \ie, $q_D\triangleq \prob{v\in \Gamma_D^{\hat{G}_j}(u)}$, where $j$ is either 1 or 2. Note that if $v$ is a $D$-hop neighbor of $u$, then $v$ is connected to some $(D-1)$-hop neighbors $i$ of $u$. Therefore, $q_D$ satisfies the following recursion:
\begin{align}\label{eq:qD}
    q_D\approx&\sum_{i\in \hat{G}_j} \prob{v \in \Gamma_1^{\hat{G}_j}(i)} \times \prob{i\in \Gamma_{D-1}^{\hat{G}_j}(u)}\nonumber\\
    \overset{(a)}{\approx}& c \int_{0}^{n^{\gamma}} n w^{-\beta} \cdot \frac{w}{n\overline{w}}\cdot wq_{D-1}dw\nonumber\\
    =&c \frac{q_{D-1}}{\overline{w}}\int_{0}^{n^{\gamma}}w^{2-\beta}dw=\frac{c n^{\gamma(3-\beta)}}{\overline{w}(3-\beta)}q_{D-1}.
\end{align}
In step $(a)$, we integrate over the degree $w$ of the $(D-1)$-hop neighbor $i$. Thus, $\prob{i \in \Gamma_{D-1}^{\hat{G}_j}(u)}$ is $w P_{D-1}$ by our definition. Further, $w/n \bar{w}$ is the probability that $v$ (with degree 1) is connected to $i$, and number of  such vertices $i$ with degree in $[w, w+dw]$ is about $ \sum_{i=1}^n \indc{w \le w_i \le w+dw} 
\to cn w^{-\beta} dw$ with $c= \left(\frac{(\beta-2)\overline{w}}{(\beta-1}\right)^{\beta-1} (\beta-1).$
By the Chung-Lu model,  $q_1\approx \frac{n^{\gamma}}{n\overline{w}}$. Iterating \prettyref{eq:qD} over $D$, it follows that 
\begin{align}
   q_{D} \approx \left(c \frac{n^{\gamma(3-\beta)}}{\overline{w}(3-\beta)}\right)^{D-1}  q_1
   \; \approx\frac{ c^{D-1} n^{\gamma\left((3-\beta)(D-1)+1\right)}}{n\overline{w}^D(3-\beta)^{D-1}}.\label{eq:intuition1}
\end{align} 
As explained in~\prettyref{sec:idea}, for the success of the $D$-hop algorithm, there are two key considerations.
On the one hand, we need to ensure that the fake pairs in $Q_1\triangleq P_1 \times P_1$ have very few $D$-hop witnesses. As such,  
we want to prevent the fake pairs in $Q_1$ from having too many common neighbors of small weight. Therefore, we require $q_D \ll 1$ 
which roughly corresponds to
$n^{\gamma((3-\beta)(D-1)+1)} \ll  n$ and is close to the condition \prettyref{eq:locallytreelikecondition} (stated later in \prettyref{thm:ChungLutheta}). 
On the other hand, we need to ensure that the true pairs in $Q_1$ have sufficiently many
$\Theta(1)$-weight $D$-hop witnesses. Indeed,  for $u \in P_1$, 
its number of common $D$-hop neighbors of $\Theta(1)$-weight is at least
\begin{align}\label{eq:intuitionof-dhoptrue}
   \abs{\{v:w_v=\Theta(1)\} \cap  \Gamma_D^{\hat{G}_1\land\hat{G}_2}(u)} \approx nq_{D} 
 \approx n^{\gamma((3-\beta)(D-1)+1)} ,
\end{align}
where the first approximation holds because there are about $\Theta(n)$ vertices with $\Theta(1)$ weight based on the power-law weight distribution. 
Therefore, under condition \prettyref{eq:theta} stated in \prettyref{thm:ChungLutheta}, which is roughly $\theta =\Omega\left( \frac{\log n}{n^{\gamma((3-\beta)(D-1)+1)}}\right)$, all the true pairs have at least $\Omega(\log n)$ low-degree $D$-hop witnesses. The above choices thus ensure that all true pairs (but no fake pairs) are matched. 

Interestingly, after matching the first slice, it triggers a cascading process,
where the new matches at one slice can be used as new seeds to match the subsequent slice by the 1-hop algorithm. To see why using 1-hop witnesses is sufficient, recall that the weight of vertices in $P_k$ satisfies
\begin{align*}
    \alpha_k \leq w_i \leq \alpha_{k-1} \Longleftrightarrow \frac{n}{\left(\frac{(\beta-1)\alpha_{k-1}}{(\beta-2)\overline{w}}\right)^{\beta-1}}-i_0\leq i \leq \frac{n}{\left(\frac{(\beta-1)\alpha_k}{(\beta-2)\overline{w}}\right)^{\beta-1}}-i_0.
\end{align*}
According to the index range of these vertices, we get that the number of vertices in $P_k$ is $\Theta\left(n\alpha_{k-1}^{1-\beta}\right)$. Since the vertices in $P_k$ and the vertices in $P_{k+1}$ are connected independently with probability at least $\frac{\alpha_k\alpha_{k+1}}{n\overline{w}}$, it follows
that, for a vertex in $P_{k+1}$, its number of 1-hop neighbors in $P_k$ is about 
\begin{align}
\label{eq:intuition-p2p1}
  n \alpha_{k-1}^{1-\beta} \times \frac{\alpha_k \alpha_{k+1}}{n \overline{w}}
  =\frac{\alpha_{k-1}^{1-\beta}\alpha_k\alpha_{k+1}}{\overline{w}}\ge\frac{\alpha_k^{3-\beta}}{8\overline{w}}.
\end{align} 
Note that  for the 1-hop algorithm to succeed, the true pairs need to have more than $\log n$ 1-hop witnesses \cite{mossel2019seeded}.
Since $2<\beta<3$, we have $\frac{\alpha_k^{3-\beta}}{8\overline{w}}>\log n$, as long as  $\alpha_k>\alpha_{k^*}\approx (\log n)^{\frac{1}{3-\beta}}$.
Therefore, assuming that the true pairs in $Q_k\triangleq P_k\times P_k$ are correctly matched, we expect that the 1-hop algorithm can correctly match the true pairs in $Q_{k+1}$ as long as $k< k^*.$

However, when $k\ge k^*$, for a vertex in $P_{k+1}$,
its number of 1-hop neighbors in $P_k$ becomes smaller
than $\log n$, and thus
the 1-hop algorithm can no longer match the vertices in $P_{k+1}$ correctly. 
Even worse,  the vertex degrees become inaccurate to distinguish the vertices with at most poly-logarithmic weights, and hence the 1-hop algorithm  can not even match the vertices slice by slice. As discussed in \prettyref{sec:idea}, we instead resort to the PGM algorithm to match a constant fraction of the rest of low-weight vertices. Note that the key to the success of the PGM is that the number of witnesses for a fake pair is no more than $2$ \cite{10.14778/2794367.2794371}. To see why this condition holds for the remaining low-weight vertices, note that the probability that a low-weight seed (with weight no larger than $\alpha_{k^*}$) becomes a 1-hop witnesses for a fake pair with weight no larger than $\alpha_{k^*}$ is at most $\left(\frac{\alpha_{k^*}\alpha_{k^*}}{n\overline{w}}\right)^2=\frac{\alpha_{k^*}^4}{n^2\overline{w}^2}$. 
Since there are at most $n$ seeds and the majority of them are low-weight, 
the number of witnesses for any fake pair with low-weights is about $\frac{\alpha_{k^*}^4}{n\overline{w}^2}\lesssim \frac{(\log n)^{\frac{4}{3-\beta}}}{n\overline{w}^2}\ll 1$. Thus, we can use the PGM algorithm with threshold $r=3$ to match a constant fraction of the low-weight vertex-pairs without errors.

Finally, the number of vertices with weight less than $\alpha_0$ is $\Theta(n)$. If most true pairs with weight less than $\alpha_0$ are matched, we can use them as new seeds to exactly match the remaining vertex-pairs in $Q_0$. 

\section{Main Results}\label{sec:result}
The following theorem provides a sufficient condition for our algorithm to correctly match a constant fraction
of nodes without any errors.
We define  $C\triangleq(2^{\beta-1}-1)\left(\frac{(\beta-2)\overline{w}}{(\beta-1)}\right)^{\beta-1}$  and  $\kappa\triangleq\frac{(1+2\delta)^22^{5-\beta}C}{(2^{3-\beta}-1)\overline{w}}$ throughout this paper.
\begin{theorem}\label{thm:ChungLutheta}
Suppose $\gamma>0$ and the positive integer $D$ are chosen such that $ \gamma \le \log_n w_{\max}$, $n^{2\gamma}=o(n)$, and
\begin{align}\label{eq:locallytreelikecondition}
     n^{\gamma((3-\beta)(D-1)+1)} \leq \frac{Cs(2^{3-\beta}-1)}{20\cdot 2^{3-\beta}}\left(\frac{Cs^2}{12 \kappa^2 \cdot\overline{w}}\right)^D \frac{n}{ (\log n)^{3-\beta} } \, .
\end{align}
If the fraction $\theta$ of seeds satisfies
\begin{align}\label{eq:theta}
    \theta\geq\frac{320\log n}{\left(\frac{Cs^2}{12\cdot\overline{w}}\right)^D n^{\gamma((3-\beta)(D-1)+1)}},
\end{align}
then for all sufficiently large $n$, Algorithm \prettyref{alg:D-hop-power-law} with $\tau_1$ in \prettyref{eq:def_tau_1} and $\tau_2(k)$ in \prettyref{eq:def_tau_2} outputs $\Theta(n)$ true pairs  and zero fake pairs with probability at least $1-n^{-1+o(1)}$.
\end{theorem}
Recall from \prettyref{eq:intuitionof-dhoptrue} that  $n^{\gamma((3-\beta)(D-1)+1)}$ is roughly the size of the $D$-hop neighborhood of a vertex (with weight around $n^\gamma$) in the first slice $P_1$. Therefore, on the one hand, \prettyref{eq:locallytreelikecondition} ensures that for two distinct vertices 
$(u,v)$ in the first slice, the intersection of their $D$-hop neighborhoods is much smaller
than the two neighborhoods, so that
the fake pairs have much fewer $D$-hop witnesses than the true pairs. 
On the other hand, \prettyref{eq:theta} ensures that the true pairs have at least $\Omega(\log n)$ $D$-hop witnesses. 

Assuming $w_{\max}=\Theta(\sqrt{n})$, if we set $D=1$ and $\gamma=\frac{1}{2}-\epsilon$ for a small constant $\epsilon>0$, then \prettyref{thm:ChungLutheta} recovers the seed requirement $n^{1/2+\epsilon}$ for the $1$-hop algorithm which is comparable to the result in \cite{10.1109/TNET.2016.2553843}. 
Surprisingly, for larger $D$, if 
we optimally choose $n^{\gamma}$ in \prettyref{eq:choice_gamma}, then the seed requirement can be
dramatically reduced to $\Omega((\log n)^{4-\beta})$, as shown by the following corollary.

\begin{corollary}[The formal version of \prettyref{thm:summary}]\label{col:leastseeds}
 Suppose 
\begin{align}
D \ge \frac{1}{3-\beta} \left( \frac{\log n}{\log (w_{\max}) } -1 \right)+1 \quad \text{ and } \quad D> \frac{4-\beta}{3-\beta}. \label{eq:cond_D}
\end{align}
Choose
\begin{align}
n^{\gamma((3-\beta)(D-1)+1)}=\frac{c n}{(\log n)^{3-\beta}}, \label{eq:choice_gamma}
\end{align}
for a sufficiently small constant $c$ so that \prettyref{eq:locallytreelikecondition} is satisfied,
and $\tau_1, \tau_2(k)$ according to \prettyref{eq:def_tau_1} and \prettyref{eq:def_tau_2}, respectively. 
If the fraction of seeds satisfies 
$$
\theta \ge \frac{C_0 \left( \log n\right)^{4-\beta}}{ n}
$$
for a sufficiently large constant $C_0$,
then
for all sufficiently large $n$, Algorithm \prettyref{alg:D-hop-power-law}  
outputs  $\Omega(n)$ true pairs and zero fake pairs, with probability at least $1-n^{-1}$.
\end{corollary}
According to~\prettyref{eq:choice_gamma}, 
we choose $\gamma$ asymptotically equal to $\frac{1}{ \left[ (3-\beta)(D-1)+1 \right]}$.
Condition \prettyref{eq:cond_D} is imposed to ensure that 
this choice
satisfies  $\gamma <1/2$ and $\gamma \le \log_n (w_{\max})$  in \prettyref{thm:ChungLutheta}.
\prettyref{thm:summary} is a special case of
Corollary \prettyref{col:leastseeds}, where
$w_{\max}=\Theta(\sqrt{n})$ so that  $ \prettyref{eq:cond_D}$ reduces to $D> \frac{4-\beta}{3-\beta}.$ 



\section{Numerical experiments}\label{sec:exp}

In this section, we conduct numerical experiments to verify our theoretical findings and the effectiveness of the PLD algorithm. For all experimental results, we calculate the accuracy rate as the median of the proportion of vertices that are correctly matched, taken over 10 independent runs.

\subsection{Choice of $D$ and $\gamma$} \label{sec:choice_gamma}
In this section, we simulate our PLD algorithm with different $D$ and $\gamma$ to investigate the impact of the two parameters. We generate the underlying parent graph $G_0$ according to the Chung-Lu model with $n=10000$, $\beta=2.5$ and $\overline{w}=10$. Then, we construct $G_1$ and $G_2$ by sampling each edge of $G_0$ twice independently with probability $s=0.8$. 
The seeds are selected such that each true pair becomes a seed  with probability $\theta$ independently.

\begin{figure}[h]
\centering
\includegraphics[width=0.7\columnwidth]{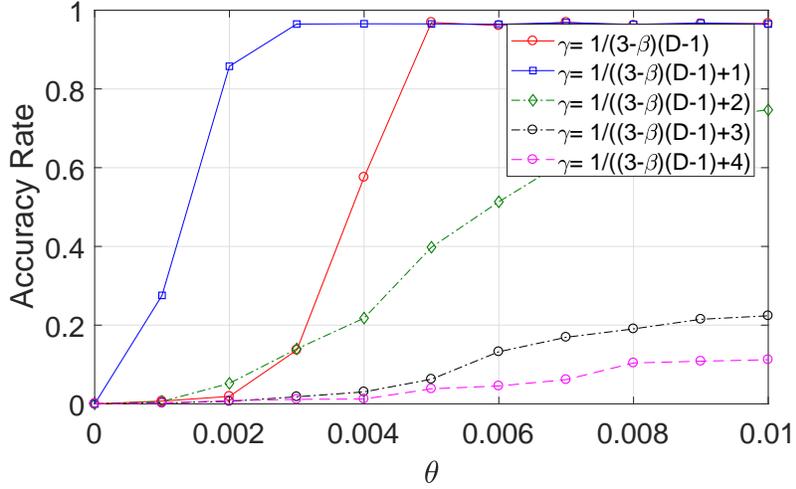}
\caption{The performance of the PLD algorithm with $D=3$ and varying $\gamma$.}
\label{fig:Choicegamma}
\end{figure}

In \prettyref{fig:Choicegamma}, we first plot the accuracy rates of our PLD algorithm with $D=3$ and different $\gamma$, when $\theta$ varies from $0$ to $0.01$. We observe that for a given accuracy rate, when $\gamma=1/\left[(3-\beta)(D-1)+1\right]$, the PLD algorithm requires the smallest number of seeds. This is consistent with the theoretical prediction in Corollary \ref{col:leastseeds}, \ie, the optimal choice of $\gamma$ approaches $1/\left[(3-\beta)(D-1)+1\right]$ as $n \to \infty$.

\begin{figure}[h]
\centering
\includegraphics[width=0.7\columnwidth]{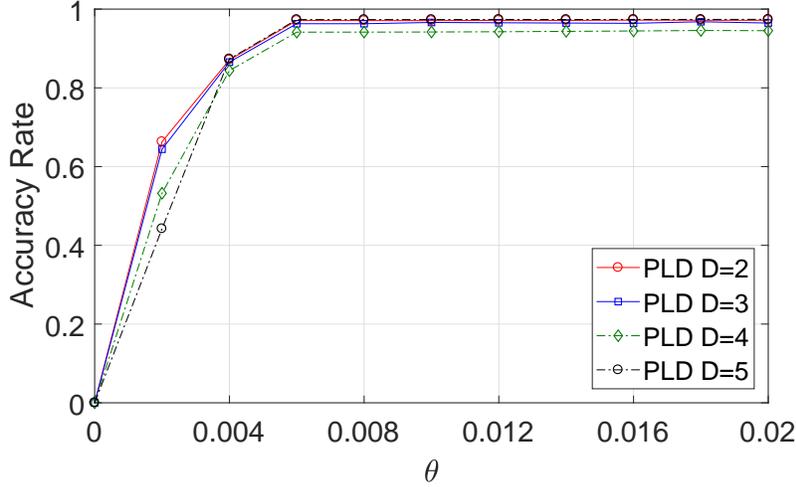}
\caption{The performance of the PLD algorithm with different $D$ and $\gamma=\frac{1}{(3-\beta)(D-1)+1}$.}
\label{fig:ChoiceD}
\end{figure}

Then, in \prettyref{fig:ChoiceD}, we plot the accuracy rates of our PLD algorithm  with different choices of $D$ by fixing $\gamma=1/[(3-\beta)(D-1)+1]$. We can see that the curves for different $D$ align well with each other, showing that the PLD algorithm with different $D$ requires a comparable number of seeds to succeed when $\gamma$ is 
optimally chosen, as suggested by Corollary \ref{col:leastseeds}.


\subsection{Performance Comparison with Synthetic Data}

\begin{figure}[h]
\centering
\includegraphics[width=0.7\columnwidth]{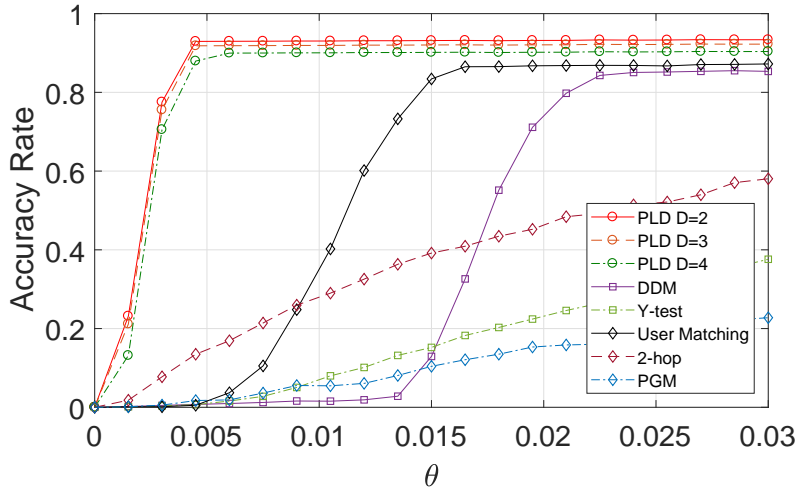}
\caption{Performance comparison of our PLD algorithm and  five other algorithms on the Chung-Lu model with different $\theta$.}
\label{fig:Compare1}
\end{figure}

For our experiments on synthetic data, we still use the graphs generated in \prettyref{sec:choice_gamma} according to the Chung-Lu model. 
Then, our PLD algorithm is simulated and compared with the 
other five state-of-the-art seeded graph matching algorithms, namely DDM \cite{10.1109/TNET.2016.2553843}, Y-test \cite{10.1007/s00453-017-0395-0}, User-Matching \cite{korula2014efficient}, 2-hop \cite{mossel2019seeded} and PGM \cite{10.14778/2794367.2794371} algorithms. 
For the PLD algorithm, we select $D=2,3,4$ and $\gamma=1/((3-\beta)(D-1)+1)$ as suggested in Corollary \ref{col:leastseeds}. 
In \prettyref{fig:Compare1}, we plot the performance comparison when $\theta$ varies from $0$ to $0.03$.  We observe that our PLD algorithm with different $D$  achieves similar performance, and it significantly outperforms  all the other algorithms.  Specifically, our PLD algorithm only requires around 50 seeds to match almost all vertices, while the User-Matching algorithm requires at least 150 seeds, and the DDM  requires at least 220 seeds. The other algorithms perform even worse. Note that roughly $5\%$ of vertices have degree at most 1 in both graphs; thus we do not expect to correctly match them. That is why the accuracy rates of our PLD algorithm saturated around $95\%$. 

Note  that the $2$-hop and PGM algorithms  have been known to work well for matching \ER graphs \cite{mossel2019seeded,10.14778/2794367.2794371}. However, we see that they are brittle to the power-law degree variations. The DDM, Y-test, and User-Matching algorithms perform slightly better. However, since they all rely on the $1$-hop witnesses, they still require a large number of seeds to succeed.

\subsection{Performance  Comparison with Real Data}\label{sec:experiment-real}

\subsubsection{Estimate Parameters for Real Graphs}\label{sec:estimate}
We see that the performance of our PLD algorithm is outstanding on synthetic graphs.
To further demonstrate the power of $D$-hops, we investigate its performance 
in matching real graphs. 
However, our algorithm based on the Chung-Lu model requires several parameters, which are unknown for real graphs.
As such, in this section, we describe our method to estimate the key model parameters before implementing our algorithm. 

First and foremost, we estimate  the power-law exponent of real graphs
by fitting them to the Chung-Lu model
using the maximum-likelihood
estimation given in \cite{clauset2009power}:
\begin{align}\label{eq:estimatebeta}
    \hat{\beta}=1+N\left[\sum_{d_i\ge d_{\min}} \text{ln}\left(\frac{d_i}{d_{\min}-1/2}\right)\right]^{-1},
\end{align}
where $d_i$ is the degree of vertex $i$, $N$  is the number of vertices with degree at least $d_{\min}$, and $d_{\min}$ is some lower bound on the vertex degrees to be specified. It is suggested in~\cite{clauset2009power} to estimate 
$d_{\min}$ using the Kolmogorov-Smirnov approach, which minimizes the maximum distance between the empirical CDF and 
the theoretical CDF of vertex degrees. More precisely, 
$$
d_{\min}=\argmin_{ d }\max_{d_i\ge d}\abs{\hat{F}_d(d_i)-F_d(d_i)},
$$
where $\hat{F}_d(x)$ is the CDF of the observed vertex degrees
with values at least $d$, and
$F(x)$ is the CDF of the power-law vertex distribution restricted
to $[d, +\infty)$. 
Numerical experiments in~\cite{clauset2009power} show $\hat{\beta}$ is accurate to $1\%$ or better if $d_{\min}$ is set to be around $6$. Thus, we fix $d_{\min}=6$ throughout our real-data experiments. 

Next, we estimate the subsampling probability $s$, which characterizes
the edge correlation between the two observed graphs. 
Let $G_j[S]$ denote the subgraph of $G_j$ induced by vertices in $S=\{i:(i,i)\in \calS\}$, where $\calS$ is the initial seed set. 
 Note that under our subsampling model,
 given an edge in one graph, it appears in the other graph  with
 probability $s.$
 Thus we  estimate the sampling probability $s$ by 
\begin{align}
\hat{s}= \frac{2\abs{E[G_1[S]\land G_2[S]]}}{\abs{E[G_1[S]]}+\abs{E[ G_2[S]]}}, \label{eq:s_est}
\end{align}
where $E[G]$ denotes the edge set of graph $G$. 

Based on $\hat{s}$, we can further estimate the average weight $\overline{w}$. Recall that $\overline{w}$ is close to the average degree under the Chung-Lu model. Thus, we estimate $\overline{w}$ 
by
$
 \frac{\overline{d}(G_1) + \overline{d}(G_2)}{2\hat{s}},
$
where $\overline{d}(G)$ is the average degree in graph $G.$
Finally, for the fraction of seeds $\theta$, if it is unknown, we can simply estimate it by $\frac{\abs{\calS}}{n}$. Note that since $w_{\max}$ will not be 
used by our algorithm, we do not need to estimate it.

Based on the estimated model parameters, we can then determine the input parameters of our PLD algorithm. Since we optimally choose $\gamma=1/((3-\beta)(D-1)+1)$, the threshold $\tau_1$ in \prettyref{eq:def_tau_1} can be simplified to 
$
\tau_1=\frac{3}{10}\left(\frac{Cs^2}{12\overline{w}}\right)^D n\theta.
$
Further, the threshold  $\tau_2(k)$ can be set according to \prettyref{eq:def_tau_2}.

\subsubsection{Facebook Friendship Networks}\label{sec:exp-facebook}
\begin{figure}[h]
\centering
\includegraphics[width=0.7\columnwidth]{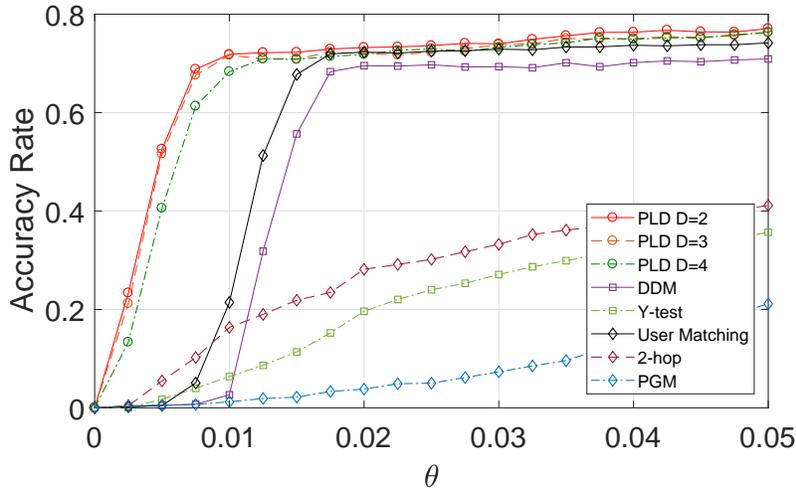}
\caption{Performance comparison of the PLD algorithm and five other algorithms applied to the Facebook networks.}
\label{fig:Compare3}
\end{figure}

We use a Facebook friendship network (provided in \cite{Traud_2012}) of 63392 students and staffs from University of Oregon as the parent graph $G_0$. There are 1633772 edges in $G_0$. The power-law exponent of the Facebook social network is estimated as 2.09 by \prettyref{eq:estimatebeta}. 
To obtain two edge-correlated subgraphs $G_1$ and $G_2$ of different sizes, 
we independently sample each edge of $G_0$ twice with probability $s=0.9$ and sample each vertex of $G_0$ twice with probability $0.8$. Then, we relabel the vertices in $G_2$ according to a random permutation $\pi:[n_2] \to [n_2]$, where $n_2$ is the number of nodes in $G_2$. Let $m$ denote the number of common vertices that appear in both $G_1$ and $G_2$. The initial seed set is constructed by including each true pair independently with probability $\theta$.
We treat $G_1$ as the public network and $G_2$ as the private network, and the goal is to de-anonymize the node identities in $G_2$ by matching $G_1$ and $G_2$. 
In \prettyref{fig:Compare3}, we show the performance of our PLD algorithm and five other algorithms, when the fraction of initial seeds $\theta$ varies from $0$ to $0.05$. We can observe that our PLD algorithm significantly outperforms the other algorithms. 

\subsubsection{Autonomous Systems Networks}\label{sec:exp-auto}
Following~\cite{pmlr-v119-fan20a}, we use the Autonomous Systems (AS) data set from \cite{snapnets} to further test the graph matching performance on power-law graphs. The data set consists of 9 graphs of Autonomous Systems peering information inferred from Oregon route-views between March 31, 2001, and May 26, 2001. Since some vertices and edges are changed over time, these nine graphs can be viewed as correlated versions of each other. The number of vertices of the 9 graphs ranges from 10,670 to 11,174 and the number of edges from 22,002 to 23,409.
We aim to match each graph to that on March 31, with vertices randomly permuted. The initial seed set is obtained by including each true pair independently with probability $\theta=0.1$.

The power-law exponent of the Autonomous Systems networks is estimated to be 2.01 according to \prettyref{eq:estimatebeta}. Note that in this experiment, the two correlated graphs
are provided  by the real data set. Thus, we further estimate 
the correlation parameter $s$ according to~\prettyref{eq:s_est}.

\begin{figure}[h]
\centering
\includegraphics[width=0.7\columnwidth]{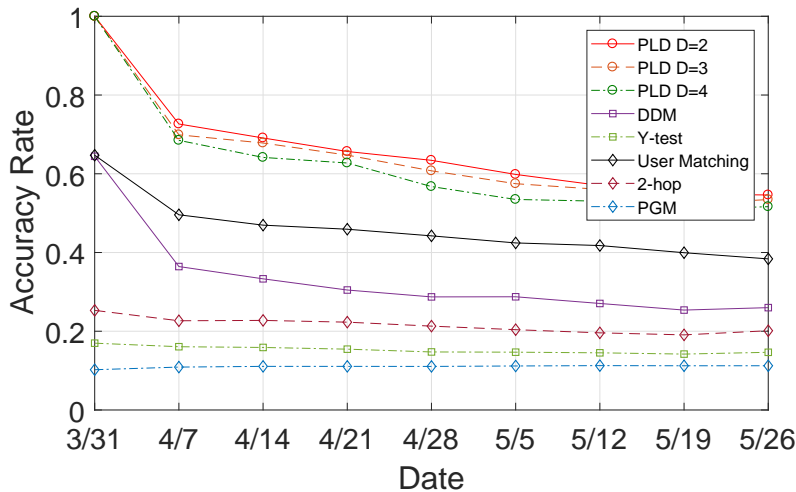}
\caption{Performance comparison of the PLD algorithm and five other algorithms applied to the Autonomous Systems  graphs  when $\theta=0.1$.}
\label{fig:Compare2}
\end{figure}

The performance comparison of the six algorithms is plotted in \prettyref{fig:Compare2} for $\theta=0.1$. We observe that our PLD algorithm again significantly outperforms the other algorithms. Note that the accuracy rates for all algorithms decay in time, because over time the graphs become less correlated with the initial one on March 31. 

\section{ Analysis}\label{sec:proof}

In this section, we present the proof for \prettyref{thm:ChungLutheta}. In \prettyref{sec:dependency}, we describe the dependency issue in our analysis and how we deal with it.
In \prettyref{sec:pf-skch-P1}, we prove that all the true pairs in 
the first slice $Q_1$ are matched error-free by the $D$-hop algorithm. 
Using the matched vertices in the previous slice as new seeds, we  show in~\prettyref{sec:pf-skch-Pk} that all the true pairs in slice $Q_k$ are matched error-free
by the $1$-hop algorithm for $2\le k \le k^*$. Further, \prettyref{sec:pf-skch-PGM} proves that using the match pairs in slice $k^*$ as new seeds, the PGM algorithm correctly matches a constant fraction of  true pairs with low weights.
Finally, in \prettyref{sec:pf-skch-P0}, we come back to  $Q_0$ and prove that using all the matched pairs as seeds, all the true pairs in $Q_0$
 are matched error-free by the $1$-hop algorithm.
\prettyref{thm:ChungLutheta} readily follows by combining these results. The proofs of auxiliary lemmas can be found in Appendix \ref{sec:proofoflmms}.

For ease of presentation, throughout the analysis, we assume without loss of generality that the true mapping $\pi$ is the identity permutation. 
We further assume $\gamma>0$ and the integer $D$ are such that $\gamma \le \log_n w_{\max}$, $n^{2\gamma}=o(n)$, and \prettyref{eq:locallytreelikecondition} holds.

\subsection{Deal with the Dependency Issues}\label{sec:dependency}
In Algorithm \ref{alg:D-hop-power-law}, we use degrees as guidance to define the imperfect slice $\hat{P}_k^{G_j}$ for $j=1,2$ 
and the induced graphs $\hat{G}_1, \hat{G}_2$. However, if we condition on the degrees, then the edges  are no longer independently generated with probability $p_{ij}$ as defined in the Chung-Lu model. To deal with this dependency issue, we construct slices based on vertex weight  that ``sandwich'' $\hat{P}_k^{G_j}$. Recall that the perfect slices defined as $P_k=\{u: w_u \in [\alpha_k,\alpha_{k-1}]\}$. By construction and the concentration of vertex degrees, 
we expect that 
$P_k \subset \hat{P}_k^{G_j}$. We also need another weight-guided slice to contain
$\hat{P}_k^{G_j}$.
Specifically, define
$$\overline{P}_k=\{u:w_u\in[(1-2\delta)\alpha_{k},(1+2\delta)\alpha_{k-1}]\},$$
where $\delta=\frac{1}{8}$. We also define $\overline{Q}_k\triangleq \overline{P}_k\times\overline{P}_k$.
The following lemma shows that  with high probability, $P_k\subset \hat{P}_k^{G_j}\subset \overline{P}_k$ and hence $Q_k\subset \hat{Q}_k\subset \overline{Q}_k$.  
Similarly, we define two different subsets of vertices that ``sandwich'' $V_j$:
$$
\underline{V}=\{u:w_u\in[0,n^{\gamma}]\} \quad \text{ and } \quad  \overline{V}=\{u:w_u\in[0,(1+2\delta)n^{\gamma}]\}.
$$
Further, let $\underline{G}_j$ and $\overline{G}_j$ denote the subgraph of $G_j$ induced by the vertex set $\underline{V}$ and $\overline{V}$, respectively, for $j=1,2$. The following lemma shows that  with high probability, 
$\underline{V} \subset V_j \subset \overline{V}$
and hence
$\underline{G}_j\subset \hat{G}_j \subset \overline{G}_j$.

\begin{lemma}\label{lmm:uinPk}
For any $0\le k \le k^*$,
\begin{align*}
\prob{Q_k\subset \hat{Q}_k\subset \overline{Q}_k }\geq 1- n^{-4+o(1)},
\end{align*}
and 
$$
\prob{Q_{\ge k^*}\subset \hat{Q}_{\ge k^*}\subset \overline{Q}_{\ge k^*} }\geq 1- n^{-4+o(1)}.
$$
For $j=1,2$,
$$
\prob{ \underline{V} \subset V_j \subset \overline{V} }=\prob{\underline{G}_j\subset\hat{G}_j\subset \overline{G}_j}\geq 1-n^{-3+o(1)}.
$$

\end{lemma}

\subsection{Match Pairs in $\hat{Q}_1$ using $D$-hop Algorithm}\label{sec:pf-skch-P1}
Recall that we give a heuristic argument of \prettyref{eq:intuitionof-dhoptrue}, showing that
for a true pair in $Q_1$, the number of common $D$-hop neighbors of $\Theta(1)$ weights is on the order of  $n^{\gamma(3-\beta)(D-1)+1}$,
by ignoring the the potential
dependency between $\hat{G}_j, \hat{Q}_1$ and graphs $G_1, G_2$.
To resolve this dependency, we crucially exploit the fact that 
with high probability $Q_1 \subset \hat{Q}_1$ and $\underline{G}_j \subset \hat{G}_j$
as shown in \prettyref{lmm:uinPk}.
In particular, we consider a true pair $(u,u)$ in ${Q}_1$ and bound its number of $\Theta(1)$-weight $D$-hop neighbors in $\underline{G}_j$.
Unfortunately, even when $\underline{G}_j \subset \hat{G}_j$, 
the  $D$-hop neighbors of $u$ in $\underline{G}_j$  may contain some vertices 
that are within the $(D-1)$-hop neighborhood of $u$ in $\hat{G}_j$, which means $\Gamma_{D}^{\underline{G}_j}
\nsubseteq \Gamma_{D}^{\hat{G}_j}$. In order to exclude such vertices, we bound the number of $\Theta(1)$-weight vertices in   $N_{D-1}^{\overline{G}_j}(u)$ from above. Fortunately,  $\big|N_{D-1}^{\overline{G}_j}(u)\big|$ is close to  $\big|\Gamma_{D-1}^{\overline{G}_j}(u)\big|$, which is on the order of $n^{\gamma(3-\beta)(D-2)+1}$ and thus is much smaller than $\big|\Gamma_{D}^{\underline{G}_j}\big|$. 
To be more precise, we have the following lemma. 

\begin{lemma}\label{lmm:boundtruedhopkslice}

Fix any vertex $u\in P_1$ and constant $c$. For all sufficiently large  $n$,
\begin{align}\label{eq:lowerboundonGamma}
   & \prob{\abs{\Gamma_D^{\underline{G}_1\land \underline{G}_2}(u)\cap \{i:w_i\le c\}}\geq \Gamma_{\min}}\geq 1-n^{-4+o(1)},
\\&\label{eq:upperboundonN}
    \prob{\abs{N_{D-1}^{\overline{G}_j}(u)\cap \{i:w_i\le c\}}\leq N_{\max}}\geq 1-n^{-4+o(1)}, \text{ for } j=1,2,
\end{align}
where $\Gamma_{\min}=\frac{1}{2}\left(\frac{C\cdot s^2}{12\cdot\overline{w}}\right)^{D}n^{\gamma((3-\beta)(D-1)+1)}$ and $N_{\max}=2c\kappa^{D}n^{\gamma((3-\beta)(D-2)+1)}$. 
\end{lemma}

To appreciate the utility of~\prettyref{lmm:boundtruedhopkslice}, note that 
under the high-probability event $\underline{G}_j \subset \hat{G}_j \subset \overline{G}_j$ for $j=1,2$, we have
$$
\Gamma_D^{\hat{G}_1}(u) \cap \Gamma_D^{\hat{G}_2}(u) 
\supset 
\Gamma_D^{\underline{G}_1\land \underline{G}_2}(u) 
\setminus \left( N_{D-1}^{\overline{G}_1}(u) \cup N_{D-1}^{\overline{G}_2}(u) \right).
$$
Therefore, combining \prettyref{eq:lowerboundonGamma} and \prettyref{eq:upperboundonN}
implies that with high probability,
\begin{align}
\abs{\Gamma_D^{\hat{G}_1}(u) \cap \Gamma_D^{\hat{G}_2}(u) \cap \{i:w_i\le c\} }
\ge \Gamma_{\min}- 2 N_{\max} \approx \Gamma_{\min},\label{eq:true_pair_d_hop}
\end{align}
where the last approximation holds because $\Gamma_{\min} \gg N_{\max}$ due to 
$2<\beta<3.$ Hence, the last display yields the desired lower bound \prettyref{eq:intuitionof-dhoptrue} to the 
number of common $D$-hop neighbors of $\Theta(1)$ weights for a true pair $(u,u) $ in $Q_1$.

Next, we adopt a similar strategy to study fake pairs. 
In particular, for a fake pair in $\hat{Q}_1$,
we bound from above its number of common $D$-hop neighbors of weights smaller than $\frac{15}{s} \log n$.\footnote{The threshold $\frac{15}{s} \log n$ is chosen such that 
$
 \{i:w_i\le \frac{15}{s}\log n\}
$
contains $\{ i: |\Gamma_1^{G_1} (i) | \le 5 \log  n , |\Gamma_1^{G_2} (i) | \le 5 \log n \}$ with high probability.}
Again, to circumvent the dependency between $\hat{G}_j,\hat{Q}_1$ and graphs $G_1,G_2$, we consider a fake pair $(u,v)$ in $\overline{Q}_1$ and bound from above its number of $\Theta(1)$-weight neighbors within the common $D$-hop neighborhood in $\overline{G}_1$
and $\overline{G}_2.$
\begin{lemma}\label{lmm:boundfakepairinP1}
  Fix any two distinct vertices  $u,v\in \overline{P}_1$. For sufficiently large $n$,
\begin{align}\label{eq:Gammauvd}
    \prob{\abs{N_D^{\overline{G}_1}(u)\cap N_D^{\overline{G}_2}(v)\cap  \{i:w_i\le \frac{15}{s}\log n\}}\leq \Psi_{\max}}
    \geq 1-n^{-4+o(1)},
\end{align}
where 
$
\Psi_{\max}=\frac{2^{3-\beta}\kappa^{2D}n^{2\gamma((3-\beta)(D-1)+1)}}{(2^{3-\beta}-1) Cn}\left(\frac{15}{s}\log n\right)^{3-\beta}+\frac{2^{\beta-2}}{2^{\beta-2}-1}\kappa^{D-1} n^{(\gamma(3-\beta)(D-2)+1)}(4+6\log n).
$ 
\end{lemma}
\begin{remark}\label{rmk:boundfakepairinP1}
To see how \prettyref{eq:Gammauvd} follows, 
note that $$N_{D}^{\overline{G}_1}(u)\cap N_{D}^{\overline{G}_2}(v)\subset \left(\Gamma_{D}^{\overline{G}_1}(u)\cup N_{D-1}(u,v)\right)\cap\left(\Gamma_{D}^{\overline{G}_2}(v)\cup N_{D-1}(u,v)\right)=\left(\Gamma_{D}^{\overline{G}_1}(u)\cap\Gamma_{D}^{\overline{G}_2}(v)\right)\cup N_{D-1}(u,v),$$
where $N_{D-1}(u,v)= N_{D-1}^{\overline{G}_1}(u)\cup N_{D-1}^{\overline{G}_2}(v)$. We have already obtained 
an upper bound to $\abs{N_{D-1}^{\overline{G}_j}}$  when proving \prettyref{eq:upperboundonN} for $j=1,2$. Thus, it remains to bound from above $\abs{\Gamma_{D}^{\overline{G}_1}(u)\cap \Gamma_{D}^{\overline{G}_2}(v)}$.
A simple yet key observation is that for a vertex $i$ of weight 1, there are two extreme cases in which $i$ becomes a common $D$-hop neighbor of $(u,v)$. 
One case is that $i$ connects to some vertex  in $\Gamma_{D-1}^{\overline{G}_1}(u)\setminus \Gamma_{D-1}^{\overline{G}_2}(v)$, and  connects to some other vertex in $\Gamma_{D-1}^{\overline{G}_2}(v)\setminus \Gamma_{D-1}^{\overline{G}_1}(u)$. It can be shown that each of these two connections happens independently with probability approximately $q_D$ and thus
the number of such common $D$-hop neighbors is about $nq_D^2$, which roughly gives rise to the first term of $\Psi_{\max}$.
The other extreme case is that $i$ is a 
$(D-1)$-hop neighbor of some common neighbor of $(u,v)$. Luckily, the common 1-hop neighborhood of $(u,v)$ is typically of a very small size and thus we can bound
from above $\abs{\Gamma_1^{\overline{G}_1}(u)\cap \Gamma_1^{\overline{G}_2}(v)}$ by approximately $\log n$. Moreover, $i$ becomes a $(D-1)$-hop neighbor of a given vertex in $\Gamma_1^{\overline{G}_1}(u)\cap \Gamma_1^{\overline{G}_2}(v)$ with probability at most $q_{D-1}$.  Thus, the number of such common $D$-hop neighbors is at most around $nq_{D-1}\log n
$,
which gives an expression close to the second term of $\Psi_{\max}$. These two extreme cases turn out to be
the dominating cases as shown in the proof of~\prettyref{lmm:boundfakepairinP1}.
\end{remark}

To see the usage of~\prettyref{lmm:boundfakepairinP1}, note that 
under the high-probability event $ \hat{G}_j \subset \overline{G}_j$ for $j=1,2$, we have
$
\Gamma_D^{\hat{G}_1}(u) \cap \Gamma_D^{\hat{G}_2}(v) 
\subset N_D^{\overline{G}_1}(u)\cap N_D^{\overline{G}_2} (v).
$
Therefore, \prettyref{eq:Gammauvd} implies that with high probability 
\begin{align}
\abs{\Gamma_D^{\hat{G}_1}(u) \cap \Gamma_D^{\hat{G}_2}(v) \cap  \{i:w_i\le \frac{15}{s}\log n \} }
\le 2 \Psi_{\max}, \label{eq:fake_pair_d_hop}
\end{align}
which yields the desired upper  bound  to the 
number of common $D$-hop neighbors of $\Theta(1)$ weights for a fake pair $(u,v) $ in $\hat{Q}_1$.

Finally, since we have $n^{\gamma(3-\beta)} \gg \log n$ and $
     n^{\gamma((3-\beta)(D-1)+1)}(\log n)^{3-\beta}= O(n)$ based on the choice in \prettyref{eq:theta}, it follows that 
     $\Gamma_{\min}> 2 \Psi_{\max}$. Moreover,
     \prettyref{eq:theta} ensures 
     that $\Gamma_{\min}\theta =\Omega(\log n)$. 
Therefore, combining \prettyref{eq:true_pair_d_hop} and \prettyref{eq:fake_pair_d_hop} implies that the true pairs in
$Q_1$ have more $D$-hop witnesses than the fake pairs in $\hat{Q}_1$. Hence, we  can use Algorithm \ref{alg:D-hop-power-law} to match pairs in $\hat{Q}_1$ correctly. More precisely, we have the following lemma. 

\begin{lemma}\label{lmm:proofP1}
Under the conditions of \prettyref{thm:ChungLutheta},
for all sufficiently large $n$,
the set of matched pairs in Step 5 of Algorithm \ref{alg:D-hop-power-law},
denoted by  $\calR_1$, contains 
 all true pairs in ${Q}_1$ and no fake pairs in $\hat{Q}_1$
  with probability at least $1-n^{-1.5+o(1)}$.
\end{lemma}

\subsection{Match Pairs in $\hat{Q}_k$ Slice by Slice using  $1$-hop Algorithm}\label{sec:pf-skch-Pk}

Given that all the true pairs in $Q_1$ are matched error-free, 
we show that all the true pairs in $Q_k$ are matched error-free
by the $1$-hop algorithm  for all $2\leq k\leq k^*$.

Note that when matching pairs in $\hat{Q}_k$, we use $\calR_{k-1}$,
the set  of matched vertices in 
$\hat{Q}_{k-1}$, as seeds. 
Suppose slice $k-1$ is successfully matched. Then,
$\calR_{k-1}$ contains all the
true pairs in $Q_{k-1}$.
Therefore, for a true pair in $Q_k$, to bound from below its number of 
$1$-hop witnesses in $\calR_{k-1}$,
it suffices to consider its number of $1$-hop common neighbors 
in $P_{k-1}$, which is on the order of $\frac{\alpha_{k-1}^{3-\beta}}{2\overline{w}}$ as we explained in \prettyref{eq:intuition-p2p1}.
This  is made precise by the following lemma.

\begin{lemma}\label{lmm:lowerbound-Pk-true}
Fix any $2\leq k\leq k^*$ and any vertex $u\in P_k$. For all sufficiently large $n$,
\begin{align}
    \prob{\abs{\Gamma_{1}^{G_1}(u)\cap \Gamma_{1}^{G_2}(u)\cap P_{k-1}}\geq  \xi_k}\geq 1-n^{-4},
\end{align}
where $\xi_k =\frac{C\alpha_{k-1}^{3-\beta}s^2}{16\overline{w}}.$
\end{lemma}

Moreover, if slice $k-1$ is successfully matched, since there is no matching error,  
on the high-probability event $\hat{P}_{k-1} \subset \overline{P}_{k-1},$
$\calR_{k-1}$ is contained by the set of true pairs in $
\overline{Q}_{k-1} \triangleq \overline{P}_{k-1}\times \overline{P}_{k-1}$.
Therefore, for a fake pair in $\hat{Q}_k$, to bound from above its
the number of $1$-hop witnesses in $\calR_{k-1}$,
it suffices to bound its number of $1$-hop common neighbors in $\overline{P}_{k-1}$,
which is done in the following lemma. Note that to resolve the potential
dependency between $\hat{Q}_k$ and graphs $G_1, G_2$,
we state the lemma for a fake pair in $\overline{Q}_k$,
which contains $\hat{Q}_k$ with high probability. 

\begin{lemma}\label{lmm:upperbound-Pk-fake}
Fix any $2\leq k\leq k^*$ and any two distinct vertices $u, v \in \overline{P}_k$,
Then for all sufficiently large $n$,
\begin{align}\label{eq:uppboundPk}
    \prob{\abs{\Gamma_{1}^{G_1}(u)\cap \Gamma_{1}^{G_2}(v)\cap \overline{P}_{k-1}}\leq \zeta_k }\geq 1-n^{-4},
\end{align}
where $\zeta_k=\frac{8(1+2\delta)^4C\alpha_{k-1}^{5-\beta}}{\overline{w}^2n}+\frac{16}{3}\log n.$
\end{lemma}
To see how~\prettyref{eq:uppboundPk} follows,
note that a vertex in $\overline{P}_{k-1}$ is a 1-hop common neighbor for the fake pair $(u,v)$ 
with probability at most on the order of $\left(\frac{\alpha_{k}\alpha_{k-1}}{n\overline{w}}\right)^2=\frac{\alpha_{k-1}^4}{4n^2\overline{w}^2}$. Since there are $\Theta(n\alpha_{k-1}^{1-\beta})$ vertices in $\overline{P}_{k-1}$, 
the number of 1-hop common neighbors in $\overline{P}_{k-1}$ is about $\frac{\alpha_{k-1}^{5-\beta}}{4n\overline{w}^2}$ on expectation. The extra term $\frac{16}{3}\log n$ in (\ref{eq:uppboundPk}) comes from the sub-exponential tail bounds when we apply concentration inequalities. 

Recall that we assume 
$n^{2\gamma}=o(n)$ and hence $\alpha_{k-1}^{3-\beta}\gg \frac{\alpha_{k-1}^{5-\beta}}{n}$ for $2\le k \le k^*$.
Moreover, $\alpha_{k-1}^{3-\beta}\ge \alpha_{k^*}^{3-\beta} \ge \frac{192\overline{w} \log n}{C s^2}$ for $2\le k \le k^*$. It then can be verified  that $\xi_k > \zeta_k$. 
Thus, we expect that the $1$-hop algorithm can match vertex pairs in $\hat{Q}_k$ correctly. 
More precisely, we have the following lemma. 
\begin{lemma}\label{lmm:proofPk}
Under the conditions of \prettyref{thm:ChungLutheta}, for all sufficiently large  $n$, with probability at least $1-n^{-1.5+o(1)}$, 
the set of matched pairs in Step 6-8 of Algorithm \ref{alg:D-hop-power-law}, 
denoted by $\calR_k$, contains
all true pairs in ${Q}_k$ and no fake pairs in $\hat{Q}_k$ for all $2\leq k\leq k^*$.
\end{lemma}

\subsection{Match Low-Weight Pairs by PGM}\label{sec:pf-skch-PGM}

We proceed to match pairs with weight  smaller than $\alpha_{k^*}$ using the PGM algorithm. As explained in \prettyref{sec:intuition}, we expect that the number of common 1-hop neighbors for any fake pair with weights smaller than $\alpha_{k^*}$ is at most $2$. Thus, even if all low-weight true pairs are provided as seeds, no fake pair will be matched by the PGM algorithm with threshold $r=3$. This is made precise by the following lemma.
\begin{lemma}\label{lmm:proofPGMfake}
Denote $\overline{P}_{\ge k^*}=\{u:w_u\in[0,(1+2\delta)\alpha_{k^*-1}]\}$. 
Fix any two distinct vertices $u, v \in \overline{P}_{\ge k^*+1}$.
Then for all sufficiently large $n,$
\begin{align}
    \prob{\abs{\Gamma_{1}^{G_1}(u)\cap \Gamma_{1}^{G_2}(v)\cap \overline{P}_{\ge k^*} }\leq 2}\geq 1-n^{-2}.
\end{align}
\end{lemma}

Although the PGM algorithm may fail to match some true pairs with very few common 1-hop neighbors, it is expected to match the true pair with at least three 1-hop witnesses. 
In particular, let us recursively define 
$$
S_{0}=P_{k^*}, \quad S_h=\{u: u\in P_{h+k^*},\ |\Gamma_1^{G_1}(u)\cap\Gamma_1^{G_2}(u) \cap S_{h-1}| \geq 3\} \quad \text{ for } h \ge 1.
$$
Note that $S_0=P_{k^*}$ has been correctly matched based on \prettyref{lmm:proofPk} in the previous step. Also, once the true pairs in $S_{h-1}$ are added into the set of matched pairs, the PGM algorithm with threshold $r=3$ can use the vertices in $S_{h-1}$ as new seeds to match vertices in $S_h$ correctly. Therefore, 
all the true pairs in $S_h$ for any $h\ge 1$ can be correctly matched.
Thus, to show the PGM matches many true pairs,
it suffices to bound from below the size of $S_h$ for $h\le h^*$, 
which is done by the following theorem.

\begin{lemma}\label{lmm:proofPGMtrue}
Let 
$ \tilde{w} \triangleq\left(\frac{192\overline{w}\ln 2}{Cs^2}\right)^{1/(3-\beta)}$.
Define $h^*$ such that $\tilde{w} \le \alpha_{k^*+h^*} < 2 \tilde{w}.$
Then for any $1 \le h\le h^*$, and all sufficiently large $n,$
\begin{align}
  \prob{\abs{S_{h}}\geq \frac{1}{2}n_{k^*+h}}\geq 1- n^{-3+o(1)}.   \label{eq:pgm_S_h}
\end{align}
\end{lemma}
The proof of \prettyref{lmm:proofPGMtrue} follows by induction. 
Assume \prettyref{eq:pgm_S_h} holds for $h-1$.
Then analogous to the intuition of \prettyref{eq:intuition-p2p1}, for any $u$ in $P_{k^*+h}$, $\expect{ \abs{\Gamma_1^{G_1}(u)\cap\Gamma_1^{G_2}(u) \cap S_{h-1}}}\approx \frac{\alpha_{k^*+h}^{3-\beta} Cs^2}{\overline{w}}  \ge 4\ln 2$. Hence, we can show that
$\prob{u \in S_h} \ge \frac{3}{4}$, which further implies \prettyref{eq:pgm_S_h} holds for $h$ by concentration.


By~\prettyref{lmm:proofPGMtrue}, the PGM matches at least half of true pairs in $P_{k^*+h^*}$.
Note that the number of vertices in $P_{k^*+h^*}$ satisfies $n_{k^*+h^*}=
C n (\alpha_{k^*+h^*-1})^{1-\beta} 
\ge Cn(\tilde{w})^{1-\beta}=\Theta(n)$, as $\tilde{w}=\Theta(1)$.
 Thus, the set of matched  pairs by the PGM contains a constant fraction of true pairs.
More precisely, we have the following lemma.

\begin{lemma}\label{lmm:proofPGM}
Under the conditions of \prettyref{thm:ChungLutheta}, for all sufficiently large  $n$, with probability at least $1-n^{-1+o(1)}$, 
the set of matched pairs in Step 10 of Algorithm \ref{alg:D-hop-power-law},
denoted by $\calR_{k^*+1}$, contains
all true pairs in $S_h$ and no fake pairs in $ 
\hat{Q}_{k^*+h}$ for all $h\ge 1$. In particular, we have $\abs{\calR_{k^*+1}}=\Theta(n)$ with probability at least $1-n^{-1+o(1)}$.
\end{lemma}

\subsection{Match Pairs in $\hat{Q}_0$ using $1$-hop Algorithm}\label{sec:pf-skch-P0}
Given that a large constant fraction of true pairs with weights smaller than $\alpha_0$ are matched error-free,
we show that all the true pairs in $Q_0$ are matched error-free
by the $1$-hop algorithm. 

When we match vertices in $\hat{Q}_0$, we use $\hat{\calR}$, the set of pairs matched in Step $5-10$ of Algorithm \ref{alg:D-hop-power-law}, as seeds.  Note that all true pairs in $Q_{k^*}$ have been proved to be matched correctly with high probability. The number of true pairs in $Q_{k^*}$ is $\Theta(n\alpha_{k^*-1}^{1-\beta})$ 
and the vertex in $P_0$ has weight larger than $n^{\gamma}$. Moreover,  a vertex in $P_0$ connects to a vertex in $P_{k^*}$ with probability at least $\frac{\alpha_{0}\alpha_{k^*}}{n\overline{w}}$. Therefore, for a true pair in $Q_0$, to bound from below its number of 
$1$-hop witnesses in $\hat{\calR}$,
it suffices to consider its number of $1$-hop common neighbors 
in $P_{k^*}$, which is about $n\alpha_{k^*-1}^{1-\beta}\times \frac{\alpha_{0}\alpha_{k^*}}{n\overline{w}} =\Theta(\alpha_{k^*}^{2-\beta}n^{\gamma})$. 
More precisely, we have the following theorem.

\begin{lemma}\label{lmm:lowerbound-P0-true}
Fix any vertex $u\in P_0$. For all sufficiently large  $n$,
\begin{align}
    \prob{\abs{\Gamma_{1}^{G_1}(u)\cap \Gamma_{1}^{G_2}(u)\cap P_{k^*}}\geq \frac{C\alpha_{k^*}^{2-\beta}\alpha_0s^2}{8\overline{w}}}\geq 1-n^{-4}.
\end{align}
\end{lemma}
We caution the reader that even though the true pair $(u,u)$ may have more $1$-hop witnesses in $Q_{k^*+1}$ than $Q_{k^*}$, we cannot consider its number of $1$-hop common neighbors in $P_{k^*+1}$, because the PGM algorithm only matches a subset of the true pairs in $Q_{k^*+1}$ and this subset is random and may incur dependency issues to the analysis.

Next we study fake pairs. Note that with high probability $\hat{\calR}$ contains no fake pair in $\bigcup_{k\ge 1}\hat{Q}_k$. 
Therefore, on the event that $\hat{P}_k \subset \overline{P}_k$ for all $k \ge 1,$
all the matched pairs in $\hat{\calR}$ is contained by the set of true pairs in $\overline{R}\times \overline{R}$, where $\overline{R}=\bigcup_{k\ge 1}\overline{P}_k=\{i:w_i\in[0,(1+2\delta)n^{\gamma}]\}$.
Therefore, for a fake pair in $\hat{Q}_0$, to bound from above its
the number of $1$-hop witnesses in $\hat{\calR}$,
it suffices to bound its number of $1$-hop common neighbors in $\overline{R}$,
which is done in the following lemma.  Again, to resolve the potential
dependency between $\hat{Q}_0$ and graphs $G_1, G_2$,
we state the lemma for a fake pair in $\overline{Q}_0$,
which contains $\hat{Q}_0$ with high probability. 

\begin{lemma}\label{lmm:upperbound-P0-fake}
Denote $\overline{R}=\{i:w_i\in[0,(1+2\delta)n^{\gamma}]\}$.
Fix any two distinct vertices $u,v\in \overline{P}_0$. For all  sufficiently large $n$, 
\begin{align}
    \prob{\abs{\Gamma_{1}^{G_1}(u)\cap \Gamma_{1}^{G_2}(v)\cap \overline{R}}\leq 4\kappa n^{\gamma(3-\beta)}s^2}\geq 1-n^{-4}, \label{eq:upperbound-P0-fake}
\end{align}
where $\kappa =\frac{(1+2\delta)^22^{5-\beta}C}{(2^{3-\beta}-1)\overline{w}}$.

\end{lemma}
To see how~\prettyref{eq:upperbound-P0-fake} follows, note that  a vertex in $P_{k}$ becomes a common 1-hop neighbor of the fake pair $(u,v)$ with probability at most  $\left(\frac{\alpha_{k}w_{\max}}{n\overline{w}}\right)^2\le\frac{\alpha_{k}^2}{n\overline{w}}$. Since there are $\Theta(n\alpha_{k}^{1-\beta})$ true pairs in $Q_{k}$, the number of common 1-hop
neighbors in $\overline{R}$ is on the order of  $\sum_{k=1}^{K}\frac{\alpha^{3-\beta}_k}{\overline{w}}= \Theta\left( n^{\gamma(3-\beta)}\right)$.

Recall that $\overline{P}_0\subset P_0\cup P_1$. Thus for any fake pair $(u,v)\in \overline{Q}_0$, the two corresponding true pairs $(u,u),(v,v)\in Q_0\cup Q_1$. 
If one of them is in $Q_1,$ then it
has already been matched in $\hat{Q}_1$ by \prettyref{lmm:proofP1}.
If one of them is in $Q_0$, since 
$\alpha_{k^*}^{2-\beta}n^{\gamma} = \Theta\left( n^{\gamma} (\log n)^{(2-\beta)/(3-\beta)} \right)\gg n^{\gamma(3-\beta)}$ in view of $2<\beta<3$,
it has more $1$-hop witnesses than the fake pair $(u,v).$
Thus, we expect that the $1$-hop algorithm can match all the true pairs in $\hat{Q}_0$ error-free. 
More precisely, we have the following lemma. 

\begin{lemma}\label{lmm:proofP0}
Under the conditions of \prettyref{thm:ChungLutheta}, for all sufficiently large  $n$, with probability at least $1-n^{-2.5}$, 
the set of matched pairs in Step 11 of Algorithm \ref{alg:D-hop-power-law}, 
denoted by $\calR_0$, contains
all true pairs in ${Q}_0$ and no fake pairs in $\hat{Q}_0$.
\end{lemma}


\subsection{Proof of \prettyref{thm:ChungLutheta}}\label{sec:pf-thm}

Due to \prettyref{lmm:proofPGM} and  $\calR_{k^*+1}\subset\calR$, the set of matched pairs by Algorithm \ref{alg:D-hop-power-law} contains $\Theta(n)$ true pairs with probability at least $1-n^{-1+o(1)}$. Combining  \prettyref{lmm:proofP1}, \prettyref{lmm:proofPk}, \prettyref{lmm:proofPGM} and \prettyref{lmm:proofP0}, $\calR$ contains no fake pairs with probability at least $1-n^{-1+o(1)}$. 


\section{Conclusion}\label{sec:conclusion}
In this paper, we propose an efficient  seeded algorithm for matching graphs with power-law degree distributions. 
Theoretically, under the Chung-Lu model with power-law exponent $2<\beta<3$ and max degree $\Theta(\sqrt{n})$, we show that as soon as $D>\frac{4-\beta}{3-\beta}$,
by optimally choosing the first slice, our algorithm correctly matches a constant fraction
of true pairs without any error with high probability, provided with only $\Omega((\log n)^{4-\beta})$ 
initial seeds. This achieves an exponential reduction in the seed size requirement,
as the previously best known result requires $n^{1/2+\epsilon}$ initial seeds. 
Empirically, numerical experiments in both synthetic and real power-law graphs further demonstrate that our algorithm significantly outperforms the state-of-the-art algorithms. These results uncover the enormous power of $D$-hops in seeded graph matching under power-law graphs. An interesting and important future direction is to further investigate the
power of $D$-hops in matching power-law graphs without seeds.

\appendix

\section{Computational Complexity Analysis}\label{sec:complexity}

We analyze the computational complexity of Algorithm \ref{alg:D-hop-power-law} in each step.

First, Algorithm \ref{alg:D-hop-power-law} checks all the vertex degrees to 
construct the subgraphs $\hat{G}_i$, ${G}'_i$ for $i=1,2$ and partition the vertices in $G_1$ and $G_2$
into slices based on vertex degrees in line 2-4 and line 9. The total time complexity
of this step is $O(n)$.

We then apply the $D$-hop algorithm in the first slice. Searching for all $D$-hop neighbors of a given vertex $u$ in the first slice takes a total of 
$O(n)$ time steps. The number of vertices in the first slice in $\Theta(n\alpha_1^{1-\beta})$. Thus,  the complexity of counting $D$-hop witnesses for  all vertices-pairs in the first slice-pair is $O(n^3\alpha_1^{2(1-\beta)})=O(n^{3-2\gamma(\beta-1)})$. Since we have shown that
with high probability, all the fake pairs have $D$-hop witnesses fewer than the threshold, we only need to sort and match at most $n$ true pairs using GMWM and hence
the complexity of the GMWM step is $O(n\log n).$

We next apply the 1-hop algorithm in the subsequent slices. We compute the number of $1$-hop witnesses via neighborhood exploration. For each matched pair in $Q_{k-1}$, we fetch its 1-hop neighbors of size $O(\alpha_{k-1})$ in $\hat{G}_1$ and $\hat{G}_2$, and then increase the number of 1-hop witnesses by 1 for $O(\alpha_{k-1}^2)$ vertex-pairs. Thus,  the total complexity of our algorithm to match vertices in $P_k$ is about $n\alpha_{k-1}^{1-\beta}\times \alpha_{k-1}^2=O(n^{1+\gamma(3-\beta)})$. Further, we match $k^*-1$ slices in line 6-8. Therefore, the total complexity is $O(n^{1+\gamma(3-\beta)}\log n)$.

Analogously,  the PGM algorithm explores the 1-hop neighbors of each matched pair. There are at most $n$ matched pair, and for each mathced pair, we increase
the number of $1$-hop witnesses by $1$ for $O(\triangle^2)$ vertex-pairs,
where $\triangle$ is the largest degree among $G_1'$ and $G_2'$.  By the definition, $\triangle$ is $O((\log n)^{\frac{1}{3-\beta}})$.
Therefore, the total complexity in line 10 is $O(n(\log n)^{\frac{2}{3-\beta}})$.

Finally, there are at most $n$ true pairs to serve as 1-hop witnesses for vertex-pairs in $\hat{Q}_0$. For any true pair $(i,i)$, the complexity of neighborhood exploration is $O(|\Gamma_1^{G_1}(i)||\Gamma_1^{G_2}(i)|)$. 
Thus, the complexity of line 11 is $\sum_{i=1}^{n}|\Gamma_1^{G_1}(i)||\Gamma_1^{G_2}(i)|=O(\sum_{i=1}^n w_i^2 )=O(n^{1+(3-\beta)/2})$ as shown in \cite[page 98]{chung2003}.

In conclusion, by summing up the complexity for each step, 
the total computational complexity of our algorithm is $O\left((n^{3-2\gamma(\beta-1)}+n\log n+n^{1+\gamma(3-\beta)}\log n+n(\log n)^{\frac{2}{3-\beta}}+n^{1+(3-\beta)/2}\right)=O\left(n^{3-2\gamma(\beta-1)}\right)$ due to $\gamma < 1/2$ and $2<\beta<3.$

\section{Postponed Proofs}

\subsection{Supporting Theorems}

\begin{theorem}\label{thm:chernoffbound} Chernoff Bound (\cite{Dubhashi2009ConcentrationOM}): Let $X=\sum_{i\in[n]}X_i$, where $X_i$, $i\in[n]$, are independent random variables taking values in $\{0,1\}$. Then, for $\eta\in [0,1]$,
\begin{align*}
\prob{X\leq (1-\eta)\expect{X}}\leq \exp\left(-\frac{{\eta}^2}{2}\expect{X}\right),\
\prob{X\geq (1+\eta)\expect{X}}\leq \exp\left(-\frac{{\eta}^2}{3}\expect{X}\right).
\end{align*}
\end{theorem}

\begin{theorem}\label{thm:bernstein} Bernstein's Inequality (\cite{Dubhashi2009ConcentrationOM}):
Let $X=\sum_{i\in[n]}X_i$, where $X_i$, $i\in[n]$, are independent random variables such that $|X_i|\leq K$ almost surely. Then, for $t> 0$, we have
\begin{align*}
\prob{X\geq \expect{X}+t}\leq \exp\left(-\frac{{t}^2}{2(\sigma^2+Kt/3)}\right),
\end{align*}
where $\sigma ^2= \sum_{i \in [n]} \var(X_i)$ is the variance of $X$. It follows then for $\rho>0$, we have
\begin{align*}
\prob{X\geq \expect{X}+\sqrt{2\sigma^2\rho}+\frac{2K\rho}{3}}\leq \exp(-\rho).
\end{align*}

The obtained estimate holds for $\prob{X\leq\expect{X}- \sqrt{2\sigma^2\rho}-\frac{2K\rho}{3}}$
too (by considering $-X$), i.e.,
\begin{align*}
\prob{X\leq \expect{X}-\sqrt{2\sigma^2\rho}-\frac{2K\rho}{3}}\leq \exp(-\rho).
\end{align*}

\end{theorem}

\begin{theorem}(\cite[Theorem~6]{yu2021graph})\label{thm:BernoulliInequality}
For $r\geq 0$, every real number $x\in (0,1)$ and $rx\leq 1$, it holds that
\begin{align*}
r\log{(1-x)}\leq \log\left(1-\frac{rx}{2} \right).
\end{align*}
\end{theorem}

\begin{theorem}(\cite[Corollary~1]{yu2021graph})\label{lmm:nkplus1issue}
Let $X$ denote a random variable such that $X\sim \Binom (n, p)$. If $n\in[n_{\min},n_{
\max}]$, then for $\lambda>0$,
\begin{align}
\prob{X\geq 2n_{\max}\alpha+\frac{4\gamma}{3}} & \leq \exp(-\gamma)
\label{eq:binom_upp_tail_2}
\end{align}
\end{theorem}

\subsection{Proof of the Main Result}\label{sec:proofoflmms}

First,  we define some notations related to graph slicing.
We count the number of vertices in the slice $P_k$ and $\overline{P}_k$. The vertices in $P_k$ satisfies
\begin{align*}
    \alpha_k\leq w_i \leq\alpha_{k-1} \Longleftrightarrow \frac{n}{\left(\frac{(\beta-1)n^{\gamma}}{(\beta-2)\overline{w}2^{k-1}}\right)^{\beta-1}}-i_0\leq i \leq \frac{n}{\left(\frac{(\beta-1)n^{\gamma}}{(\beta-2)\overline{w}2^{k}}\right)^{\beta-1}}-i_0.
\end{align*}

According to the index range of the vertices, we define $n_k$ to be the difference between the two bounds. To be more precise, 
\begin{align}\label{eq:Pknumber}
   n_k\triangleq Cn\alpha_{k-1}^{1-\beta},
\end{align}
where $C$ throughout this paper denotes $(2^{\beta-1}-1)\left(\frac{(\beta-2)\overline{w}}{(\beta-1)}\right)^{\beta-1}$. Moreover,
we have that 
\begin{align}
 n_k\leq \abs{P_k}\leq n_k+1\leq\frac{11}{10}n_k.
\end{align}

Similarly, the vertices in $\overline{P}_k$ satisfies
\begin{align*}
    (1-2\delta)\alpha_k\leq w_i \leq(1+2\delta)\alpha_{k-1}.
\end{align*}
Thus, 
\begin{align}
    \abs{\overline{P}_k}
    \le&\left(2^{\beta-1}(1+2\delta)^{\beta-1}-(1-2\delta)^{\beta-1}\right)\frac{n_k}{2^{\beta-1}-1}+1\nonumber\\
    \overset{(a)}{\le}& \frac{\left(\frac{5}{2}\right)^{\beta-1}-\left(\frac{3}{4}\right)^{\beta-1}}{2^{\beta-1}-1}n_k+1
    \le 2n_k,\label{eq:numberPkbar}
\end{align}
where  $(a)$ follows from  $\delta=\frac{1}{8}$.

The number of perfect slices, denoted by $K$, is
\begin{align*}
    \log_2 \left( n^{\gamma} \right)\leq K\leq 1+\log_2 \left( n^{\gamma} \right).
\end{align*}

\subsubsection{Proof of \prettyref{lmm:uinPk}}

First, we prove $P_k\subset \hat{P}_k^{G_j}$ with high probability 
for $0\le k\le k^*$ and $j=1,2$.
Fix any vertex $u$ in $P_k$. 
It suffices to show 
with high probability $u \in \hat{P}_k^{G_j}$. 
Note that any vertex $v$ connects to $u$ in $G_j$ independently with probability $p_{uv}s$, where $j=1,2$
and $p_{uv}=\frac{w_u w_v}{n\overline{w}}$.
Thus 
$$
\expect{\abs{\Gamma_1^{G_j}(u)}}=\sum_{v\in G_j}p_{uv}s=w_u s. 
$$
Note that $\alpha_{k}\leq w_u\leq \alpha_{k-1}$ and $\alpha_k\ge \alpha_{k^*}\ge \left(\frac{192\overline{w}\log n}{Cs^2}\right)^{\frac{1}{3-\beta}}\ge \frac{20\log n}{\delta^2s}$ 
for the choice of $k^*$ in \prettyref{eq:choice_k_star} and sufficiently large $n$, in view of $2<\beta<3.$ Then, applying the Chernoff Bound in \prettyref{thm:chernoffbound} with $\eta=\delta$ yields 
 \begin{align*}
     \prob{\abs{\Gamma_1^{G_j}(u)}\geq (1+\delta)\alpha_{k-1}s}\leq \exp\left(-\delta^2\frac{\alpha_{k-1}s}{3}\right)\leq n^{-5},
 \end{align*}
 and
  \begin{align*}
     \prob{\abs{\Gamma_1^{G_j}(u)}\leq (1-\delta)\alpha_{k}s}\leq \exp\left(-\delta^2\frac{\alpha_{k}s}{2}\right)\leq n^{-5}.
 \end{align*}
Combining the last two displayed equation yields that
$$
\prob{u\notin \hat{P}_k^{G_j}} \le 2n^{-5}.
$$
Taking an union bound over $u$ gives
 \begin{align}
\prob{P_k\subset \hat{P}_k^{G_j}}\geq 1-\sum_{u\in P_k}\prob{u\notin \hat{P}_k^{G_j}}\geq  1- n^{-4+o(1)}. \label{eq:P_k_subset}
\end{align}

Next we show that  $P_{\ge k^*}\subset \hat{P}_{\ge k^*}^{G_j}$ with high probability. 
Fix any vertex $u\in P_k$ with $k\ge k^*$. Take a vertex $v \in P_{k^*}$ with $w_v=\alpha_{k^*-1}$. Since $w_u \le w_v$, we have $\abs{\Gamma_1^{G_j}(u)}\overset{s.t.}{\le}\abs{\Gamma_1^{G_j}(v)}$.
Therefore, 
 \begin{align*}
  \prob{ u \notin \hat{P}_{\ge k^*}^{G_j}}
  =\prob{\abs{\Gamma_1^{G_j}(u)}\geq (1+\delta)\alpha_{k^*-1}s}\le \prob{\abs{\Gamma_1^{G_j}(v)}\geq (1+\delta)\alpha_{k^*-1}s}\leq  n^{-5},
 \end{align*}
Taking a union bound over $u$ gives 
\begin{align}
\prob{P_{\ge k^*}\subset \hat{P}_{\ge k^*}^{G_j}} \geq 1-n^{-4+o(1)}. \label{eq:P_k_subset_2}
\end{align}
 
Second, we prove that for $0\le k\le k^*$, with high probability $\hat{P}_k\subset \overline{P}_k$, or equivalently, 
$[n]\backslash  \overline{P}_k \subset [n]\backslash \hat{P}_k$, 
Fix any vertex $u$ with $w_u>(1+2\delta)\alpha_{k-1}$, applying the Chernoff Bound with $\eta=\frac{\delta}{1+2\delta}$ yields
 \begin{align}
  \prob{\abs{\Gamma_1^{G_j}(u)}\leq (1+\delta)\alpha_{k-1}s}\leq   \exp\left(-\delta^2\frac{\alpha_{k-1}s}{2(1+2\delta)}\right) \leq n^{-5}. \label{eq:Gamma_bound_1}
 \end{align}
 
 For any vertex $u$ with $w_u<(1-2\delta)\alpha_{k}$, applying the Chernoff Bound with $\eta=\frac{\delta}{1-2\delta}$ yields
 \begin{align}
   \prob{\abs{\Gamma_1^{G_j}(u)}\geq (1-\delta)\alpha_{k}s}\leq   \exp\left(-\delta^2\frac{\alpha_{k}s}{3(1-2\delta)}\right)\leq n^{-5}. \label{eq:Gamma_bound_2}
 \end{align}
 
 Thus, we have
\begin{align}
\prob{\hat{P}^{G_j}_k\subset \overline{P}_k}
= \prob{[n]\backslash \overline{P}_k \subset [n]\backslash \hat{P}^{G_j}_k}
\geq 1-\sum_{u\notin \overline{P}_k}\prob{u\in \hat{P}^{G_j}_k}\ge 1-n^{-4}, \label{eq:P_k_overline}
\end{align}
where the last inequality holds by combining~\prettyref{eq:Gamma_bound_1} and~\prettyref{eq:Gamma_bound_2}.
Moreover, 
\begin{align}
\prob{\hat{P}^{G_j}_{\ge k^*}\subset \overline{P}_{\ge k^*}}
=\prob{[n] \backslash \overline{P}_{\ge k^*} \subset [n] \backslash \hat{P}^{G_j}_{\ge k^*}} 
\ge 1- \sum_{u:w_u>(1+2\delta)\alpha_{k^*-1}}\prob{u\in \hat{P}_{\ge k^*}^{G_j}} \geq 1-n^{-4}, \label{eq:P_K_overline_2}
\end{align}
where the last inequality holds
by~\prettyref{eq:Gamma_bound_1}.
 
Then, combining~\prettyref{eq:P_k_subset} and~\prettyref{eq:P_k_overline} with the union bound yields that $\prob{Q_k\subset \hat{Q}_k\subset \overline{Q}_k }\geq 1- n^{-4+o(1)}$ for $0 \le k \le k^*$.
Similarly, combining~\prettyref{eq:P_k_subset_2} and~\prettyref{eq:P_K_overline_2} with a union bound
yields that
$
\prob{Q_{\ge k^*}\subset \hat{Q}_{\ge k^*}\subset \overline{Q}_{\ge k^*} }\geq 1- n^{-4+o(1)}.
$

Finally, since $\underline{V}=\bigcup_{k\ge 1}P_k$, $V=\bigcup_{k\ge 1}\hat{P}^{G_j}_k$ and $\overline{V}=\bigcup_{k\ge 1}\overline{P}_k$,
combining~\prettyref{eq:P_k_subset},~\prettyref{eq:P_k_subset_2},
~\prettyref{eq:P_k_overline}, and~\prettyref{eq:P_K_overline_2}
with the union bound, we have
$$
  \prob{\underline{G}_j\subset\hat{G}_j\subset \overline{G}_j}=  \prob{\underline{V} \subset V_j \subset \overline{V} }
\geq 1-n^{-3+o(1)}.
$$

\subsubsection{Proof of \prettyref{lmm:boundtruedhopkslice}}\label{sec:pf-lmm1}

Note that $\underline{G}_1\land\underline{G}_2$, $\overline{G}_1,$ and $\overline{G}_2$ are graphs that are edge-sampled from $G_0$ with probability $s^2$, $s$, $s$, respectively. Thus, we let $G$ denote a graph obtained by sampling each edge of $G_0$ independently with probability $t=\Theta(1)$ and $\overline{G}$ denote a subgraph of $G$ induced by the vertex set $\overline{V}=\{u:w_u\in[0,(1+2\delta)n^{\gamma}]\}$.  Fix a vertex $u\in P_1$,  we first study its number of $d$-hop neighbors in each slice in $\overline{G}$. Then, we can arrive at \prettyref{lmm:boundtruedhopkslice} by selecting the corresponding parameters. To be more precise, we define $\Gamma_{d,k}^{\overline{G}}(u)=\Gamma_{d}^{\overline{G}}(u)\cap \overline{P}_k$ and $N_{d,k}^{\overline{G}}(u)=\bigcup_{1\le j \le d} \Gamma_{j,k}^{\overline{G}}(u)$. We bound $\Gamma_{d,k}^{\overline{G}}(u)$ and $N_{d,k}^{\overline{G}}(u)$ by the following lemma.
\begin{lemma}\label{lmm:Omega}
Fix any vertex $u\in P_1$, and let $\Omega_{d}$ denote the event such that the followings hold simultaneously for $k=1, \ldots, K$:
\begin{align}
\abs{\Gamma_{d,k}^{\overline{G}}(u)} &\ge  2^{(k-1)(\beta-2)}\left(\frac{(1-2\delta)^2C\cdot t}{12\cdot\overline{w}}\right)^{d} n^{\gamma(3-\beta)d}\triangleq \Gamma_{\min}(d,k),\label{eq:Gammadku_lowerbound}\\
    \abs{\Gamma_{d,k}^{\overline{G}}(u)} & \le 2^{(k-1)(\beta-2)}\kappa^{d}n^{\gamma(3-\beta)d}\triangleq \Gamma_{\max}(d,k),\label{eq:Gammadku_upperbound}\\
 \abs{N_{d,k}^{\overline{G}}(u)} &\leq  2^{(k-1)(\beta-2)+1}\kappa^{d}n^{\gamma(3-\beta)d},\label{eq:Ndku}
\end{align}
where $\kappa=\frac{(1+2\delta)^22^{5-\beta}C}{(2^{3-\beta}-1)\overline{w}}$. 
Suppose $\gamma$ and $D$ are  chosen such that 
condition \prettyref{eq:locallytreelikecondition} holds.
Then,  for all $1\le d \le D$ and sufficiently large  $n$,
\begin{align}\label{eq:Omega}
\prob{\Omega_d}\geq 1-(4^{d}-1)n^{-4}.
\end{align}
\end{lemma}

\begin{remark}
The intuition behind~\prettyref{lmm:Omega} is as follows. Recall that $q_d$, the probability that a vertex of $\Theta(1)$ weight lies in the $d$-hop neighborhood of a vertex in the first slice, is on the order of $n^{\gamma[(3-\beta)(d-1)+1]-1}$ in view of \prettyref{eq:intuition1}. Note that the weight of vertices in $P_k$ is about $\alpha_k$, and  the size of $P_k$ is $\Theta(n\alpha_{k-1}^{1-\beta})$. Thus, the expected number of  vertices in $P_k$ that are $d$-hop neighbors of a given vertex in the first slice is roughly $nq_d\alpha_{k-1}^{2-\beta}\approx 2^{(k-1)(\beta-2)}n^{\gamma(3-\beta)d}$.
Hence, we expect~\prettyref{eq:Gammadku_lowerbound}-\prettyref{eq:Ndku} to hold with high probability by concentration.
\end{remark}


Before proving~\prettyref{lmm:Omega}, we first show how
to apply~\prettyref{lmm:Omega} to prove~\prettyref{lmm:boundtruedhopkslice}. 
By setting $\delta=0$ and $t=s^2$, we have $\overline{G}=\underline{G}_1\land \underline{G}_2$. Thus,  \prettyref{eq:Gammadku_lowerbound} with $k= \lceil\log_2(n^{\gamma})\rceil$ and $d=D$  leads to the desired conclusion \prettyref{eq:lowerboundonGamma}.
Moreover, there are at most $c$ slices in $\{i:w_i\le c\}$. By setting $\delta=\frac{1}{8}$, $d=D-1$, $\overline{G}=\overline{G}_j$ (\ie, $t=s$), \prettyref{eq:Ndku} with $\log_2(n^{\gamma}/c) \le k\le K\le \log_2 (n^{\gamma})+1$, we have
\begin{align*}
    \sum_{k=\lfloor \log_2(n^{\gamma}/c) \rfloor}^K 2^{(k-1)(\beta-2)+1}\kappa^{D-1}n^{\gamma(3-\beta)(D-1)} \leq 2c\kappa^{D-1}n^{\gamma((3-\beta)(D-2)+1)}=N_{\max},
\end{align*}
where $N_{\max}$ is given in \prettyref{eq:upperboundonN}. Thus, we prove  the desired conclusion \prettyref{eq:upperboundonN}.

We then present the proof of \prettyref{lmm:Omega}.
\begin{proof}[Proof of \prettyref{lmm:Omega}]
Fix a vertex $u$ in $P_1$, we study its $d$-hop neighborhood in $\overline{G}$ from $d=1$. 
\paragraph{For $d=1$:} For each vertex $i\in \overline{P}_k$, define an indicator variable 
$$
x_i^k=\indc{i\in \Gamma_1^{\overline{G}}(u)}.
$$
In other words, $x_i^k=1$ if $i$ is connected to $u$  in $\overline{G}$, and $x_i^k=0$ otherwise.  Since  $u\in P_1$, it follows that 
\begin{align*}
 p_{\min}^k= (1-2\delta)\frac{\alpha_k\alpha_1}{n\overline{w}}t \le\prob{x_i^k=1}\le (1+2\delta)\frac{\alpha_{k-1}\alpha_0}{n\overline{w}}t=p_{\max}^k.
\end{align*}

Then, we have $\abs{\Gamma_{1,k}^{\overline{G}}(u)}=\sum_{i\in \overline{P}_k}x_i^k$ and $x_i^k$'s are independent. Recall that
$n_k=Cn\alpha^{1-\beta}_{k-1}$ in view of \prettyref{eq:Pknumber} and $ n_k\le\abs{\overline{P}_k}\le 2n_k$ in view of \prettyref{eq:numberPkbar}. Thus
\begin{align*}
n_k p_{\min}^k  &= (1-2\delta) C \frac{\alpha_{k-1}^{2-\beta} \alpha_1}{ 2\overline{w} } t 
=(1-2\delta) C \frac{n^{\gamma (3-\beta) }}{4 \cdot 2^{(k-1)(2-\beta)} \overline{w} } t,  \\
n_k p_{\max}^k &=(1+2\delta) C \frac{\alpha_{k-1}^{2-\beta} \alpha_0}{ \overline{w} } t 
=(1+2\delta) C \frac{n^{\gamma (3-\beta) }}{ 2^{(k-1)(2-\beta)} \overline{w} } t.
\end{align*}

Hence, applying Chernoff Bound in \prettyref{thm:chernoffbound} with $\eta=\frac{1}{2}$ yields that
\begin{align*}
&\prob{\abs{\Gamma_{1,k}^{\overline{G}}(u)}\leq (1-2\delta)\frac{Cn^{\gamma(3-\beta)}t}{8\cdot 2^{(k-1)(2-\beta)}\overline{w}}}
\le \prob{\Binom\left(n_{k},p_{\min}^k\right)
\le\frac{1}{2}n_kp_{\min}^k} \overset{(a)}{\le} n^{-4},
\end{align*}
\begin{align*}
&\prob{\abs{\Gamma_{1,k}^{\overline{G}}(u)}\geq (1+2\delta)\frac{3Cn^{\gamma(3-\beta)}t}{  2^{(k-1)(2-\beta)}\overline{w}}}
\le \prob{\Binom\left(2n_{k},p_{\max}^k\right)
\le 3n_kp_{\max}^k} \overset{(b)}{\le} n^{-4},
\end{align*}
where $(a)$ and $(b)$ hold because
$
n_kp_{\max}^k\ge n_kp_{\min}^k  \ge(1-2\delta)\frac{Cn^{\gamma(3-\beta)}t}{4\cdot \overline{w}}\ge 108\log n
$ for sufficiently large $n$.

We also have $\prob{\abs{N_{1,k}^{\overline{G}}(u)}\ge 3n_kp_{\max}^k}\le n^{-4}$ due to $N_{1,k}^{\overline{G}}(u)=\Gamma_{1,k}^{\overline{G}}(u)$. 
Finally, taking the union bound leads to $\prettyref{eq:Omega}$ for $d=1$.

\paragraph{For $2\le d\le D$:} We first count the $d$-hop neighbors conditional on the $(d-1)$-hop neighborhood of $u$ such that $\Omega_{d-1}$ holds.  
The high-level idea is as follows. After the conditioning, 
every vertex $i$ outside the $(d-1)$-hop neighborhood of $u$ will become
a $d$-hop neighbor by connecting to at least one of the 
$(d-1)$-hop neighbors $v$ of $u$. These edge connections are
still independently generated 
across different $v$ and $i$ according to the Chung-Lu model.


We first bound $\abs{\Gamma_{d,k}^{\overline{G}}(u)}$ from below.  For each vertex $i\in \overline{P}_k \setminus \left(N_{d-1,k}^{\overline{G}}(u)\right)\triangleq P_k'$, define an indicator variable $$
y_i^k=\indc{\exists v\in \Gamma_{d-1}^{\overline{G}}(u):\; i\in \Gamma_1^{\overline{G}}(v)}.
$$
In other words, $y_i^k=1$ if $i$ is connected to at least one $(d-1)$-hop neighbor of $u$ in $\overline{G}$, and $y_i^k=0$ otherwise. 
Thus, we have
$
\abs{\Gamma_{d,k}^{\overline{G}}(u)} = \sum_{i \in P_k'} y_i^k,
$ and $y_i^k$'s are independent across different
$i$ conditional on $\Omega_{d-1}$.

Note that $\Gamma_{d-1,1}^{\overline{G}}(u)\subset \Gamma_{d-1}^{\overline{G}}(u)$. Thus, 
we can bound $\prob{y_i^k=1|\Omega_{d-1}}$ from below by considering
the possible edge connections between $i$ and vertices in 
$\Gamma_{d-1,1}^{\overline{G}}(u)$. More precisely, we get that 
\begin{align*}
    \prob{y_i^k=1\mid \Omega_{d-1}}\ge&\prob{\exists v\in \Gamma_{d-1,1}^{\overline{G}}(u): i\in \Gamma_1^{\overline{G}}(v)\mid \Omega_{d-1} }\\
  \overset{(a)}{\ge} & 1-\left(1-p_{vi}\right)^{\Gamma_{\min}(d-1,1)} \\
    \overset{}{\ge}& 1-\left(1-(1-2\delta)^2\frac{\alpha_{k}\alpha_{1}}{n\overline{w}}t\right)^{\Gamma_{\min}(d-1,1)}\\
    \overset{(b)}{\ge}&\frac{(1-2\delta)^2}{2}  \Gamma_{\min}(d-1,1) \frac{\alpha_{k}\alpha_{1}t}{n\overline{w}}\\
    \overset{}{=}
    &\frac{3}{2^{k}Cn}\left(\frac{(1-2\delta)^2C\cdot t}{12\cdot\overline{w}}\right)^{d} n^{\gamma((3-\beta)(d-1)+2)}\triangleq p_{\min}^{k,d}.
\end{align*}
where $(a)$ holds because $\left\{i\notin \Gamma_1^{\overline{G}}(v)\right\}$ are independent across $v$; $(b)$ follows from \prettyref{thm:BernoulliInequality}.

Now, to bound $\abs{\Gamma_{d,k}^{\overline{G}}(u)}$ from below, we also need
a lower bound to $|P_k'|$, or equivalently an upper bound to 
$\abs{N_{d-1,k}^{\overline{G}}(u)}$.
Since we have conditioned on the $(d-1)$-hop neighborhood of
$u$ such that event $\Omega_{d-1}$ holds. It follows from~\prettyref{eq:Ndku} that 
\begin{align*}
   \abs{N_{d-1,k}^{\overline{G}}(u)}\leq & 2^{(k-1)(\beta-2)+1}\kappa^{d-1}n^{\gamma(3-\beta)(d-1)}\\
   = &2\kappa^{d-1}n^{\gamma((3-\beta)(d-2)+1)}\alpha_{k-1}^{1-\beta}\\
    \overset{(a)}{\le} &\frac{C}{10}n \alpha_{k-1}^{1-\beta}\leq\frac{1}{9}n_k,
\end{align*}
where $(a)$ holds due to the condition \prettyref{eq:locallytreelikecondition}. Thus, we have $\abs{P_k'}\ge \abs{P_k}-\abs{N_{d-1,k}^{\overline{G}}(u)}\ge \frac{8}{9}n_k$. 

Note that for sufficiently large $n$,
$$
\frac{8}{9}n_kp_{\min}^{k,d}=\frac{4}{3\cdot2^{(k-1)(2-\beta)}}\left(\frac{(1-2\delta)^2Ct}{12\cdot\overline{w}}\right)^{d}n^{\gamma(3-\beta)d}= \frac{4}{3}\Gamma_{\min}(d,k)\geq 128\log n.
$$
Thus, we apply the Chernoff Bound in \prettyref{thm:chernoffbound} with $\eta=\frac{1}{4}$ and get
\begin{align*}
    &\prob{\abs{\Gamma_{d,k}^{\overline{G}}(u)}\leq \Gamma_{\min}(d,k)\mid\Omega_{d-1}}\le \prob{\Binom\left(\frac{8}{9}n_k,p_{\min}^{k,d}\right)\leq \Gamma_{\min}(d,k)\mid\Omega_{d-1}}
    \leq n^{-4}.
\end{align*}

Next, we bound $\abs{\Gamma_{d,k}^{\overline{G}}(u)}$ from above. 
To this end, we bound $\prob{\overline{y}_i^k=1|\Omega_{d-1}}$ from above and get
\begin{align}\label{eq:pmaxkd}
    \prob{\overline{y}_i^k=1|\Omega_{d-1}} \overset{(a)}{\le}&\sum_{l=1}^{K}\prob{\exists j\in \Gamma_{d-1,l}^{\overline{G}}(u): i\in \Gamma_1^{\overline{G}}(j) \mid \Omega_{d-1}}\nonumber\\
    \overset{(b)}{\le} &(1+2\delta)^2\sum_{l=1}^{K} \Gamma_{\max}(d-1,l) \frac{\alpha_{k-1}\alpha_{l-1}}{n\overline{w}}\nonumber\\
    =&(1+2\delta)^2 \frac{\kappa^{d-1}n^{\gamma((3-\beta)(d-1)+2)}}{2^{k-1}n\overline{w}} \sum_{l=1}^{K}2^{(l-1)(\beta-3)}\nonumber \\    \overset{}{\le}
    &\frac{\kappa^{d}n^{\gamma((3-\beta)(d-1)+2)}}{2^{k+1}Cn} \triangleq p_{\max}^{k,d},
\end{align}
where  $(a)$ follow from the union bound; $(b)$ holds due to the union bound and event $\Omega_{d-1}$; $(b)$ follows from $(1+x)^r\geq 1+rx$ for every integer $r \geq 0$ and every real number $x \geq -2$; 
and the last inequality follows from the definition of 
$\kappa=\frac{(1+2\delta)^22^{5-\beta}C}{(2^{3-\beta}-1)\overline{w}}$.

Also, note that $P_k'\subset \overline{P}_k$ and
thus $|P_k'| \le |\overline{P}_k| \le 2n_k $.
For sufficiently large $n$, we have
$$2n_kp_{\max}^{k,d}=2^{(k-1)(\beta-2)-1}{\kappa^{d}n^{\gamma(3-\beta)d}} =\frac{1}{2}\Gamma_{\max}(d,k).
$$

Hence, applying Chernoff Bound in \prettyref{thm:chernoffbound} with $\eta=1$ yields that
\begin{align*}
    &\prob{\abs{\Gamma_{d,k}^{\overline{G}}(u)}\geq \Gamma_{\max}(d,k)\mid \Omega_{d-1}}\le \prob{\Binom\left(2n_k,p_{\max}^{k,d}\right)\geq  \Gamma_{\max}(d,k)}\leq n^{-4}.
\end{align*}

\paragraph{Induction:} Finally, we prove \prettyref{eq:Omega} by induction.


For $d=1$, we have proved that \prettyref{eq:Omega} holds.
Suppose that \prettyref{eq:Omega} holds for $d-1$. Then we have
\begin{align}\label{eq:lowerboundonGammak}
    &\prob{\abs{\Gamma_{d,k}^{\overline{G}}(u)}\leq \Gamma_{\min}(d,k)}\le \prob{\Omega_{d-1}^c}+ \prob{\abs{\Gamma_{d,k}^{\overline{G}}(u)}\leq \Gamma_{\min}(d,k)\mid\Omega_{d-1}}\prob{\Omega_{d-1}} \leq 4^{d-1} \cdot n^{-4}.
\end{align}

Similarly, we get
\begin{align}\label{eq:upboundonGammak}
   & \prob{\abs{\Gamma_{d,k}^{\overline{G}}(u)}\geq \Gamma_{\max}(d,k)}
   \leq 4^{d-1}\cdot n^{-4}
\end{align}

Since $\abs{N_{d,k}^{\overline{G}}(u)}=\abs{N_{d-1,k}^{\overline{G}}(u)}+\abs{\Gamma_{d,k}^{\overline{G}}(u)}$, we  take an union bound and have
\begin{align}\label{eq:upperboundonNk}
    \prob{\abs{N_{d,k}^{\overline{G}}(u)}\geq 2^{{(k-1)(\beta-2)}+1}\kappa^dn^{\gamma(3-\beta)d}}\leq &(4^{d-1}-1) \cdot n^{-4}+4^{d-1} \cdot  n^{-4}=(2\cdot 4^{d-1}-1)n^{-4}.
\end{align}

Combining \prettyref{eq:lowerboundonGammak}, \prettyref{eq:upboundonGammak} and \prettyref{eq:upperboundonNk} with an union bound, we prove that \prettyref{eq:Omega} holds for any $1
\le k\le K$ and $1\le d\le D$.

\end{proof}

\subsubsection{Proof of \prettyref{lmm:boundfakepairinP1}}\label{sec:pf-lmm2}

Note that 
\begin{align}
    N_{D,k}^{\overline{G}_1}(u)\cap N_{D,k}^{\overline{G}_2}(v)
    & \subset \left(\Gamma_{D,k}^{\overline{G}_1}(u)\cup N_{D-1,k}(u,v)\right)\cap\left(\Gamma_{D,k}^{\overline{G}_2}(v)\cup N_{D-1,k}(u,v)\right)\nonumber \\
    & =\left(\Gamma_{D,k}^{\overline{G}_1}(u)\cap\Gamma_{D,k}^{\overline{G}_2}(v)\right)\cup N_{D-1,k}(u,v),
    \label{eq:set_inclusion_D}
\end{align}
where $N_{D-1,k}(u,v)= N_{D-1,k}^{\overline{G}_1}(u)\cup N_{D-1,k}^{\overline{G}_2}(v)$.
Since we have already obtained the upper bounds of $\abs{N_{D-1,k}^{G_1}(u)}$ and $\abs{N_{D-1,k}^{G_2}(v)}$ by \prettyref{lmm:Omega} by letting $\overline{G}$ to be either $\overline{G}_1$ or $\overline{G}_2$, it remains to bound
from above $\abs{\Gamma_{D,k}^{\overline{G}_1}(u)\cap \Gamma_{D,k}^{\overline{G}_2}(v)}$, which is done in the following lemma.  

\begin{lemma}\label{lmm:Gammauvk}
Suppose $\gamma$ and $D$ are  chosen such that 
condition \prettyref{eq:locallytreelikecondition} holds. Fix any two distinct vertices $u,v\in \overline{P}_1$,  for all $1\le d \le D$, $k=1, \ldots, K$, and sufficiently large $n$,
\begin{align}\label{eq:Gammauvk}
    \prob{\abs{\Gamma_{d,k}^{\overline{G}_1}(u)\cap \Gamma_{d,k}^{\overline{G}_2}(v)}\le\Psi(d,k)}\geq 1-\frac{2\cdot 4^d}{3}\cdot n^{-4},
\end{align}
where 
$$ 
\Psi(d,k)=  \frac{\kappa^2\Gamma_{\max}^2(d-1,1)n^{\gamma(5-\beta)}}{ 2^{(k-1)(3-\beta)}Cn}+\frac{6\Gamma_{\max}(d-1,1)\log  n}{2^{(k-1)(2-\beta)}}
$$
with $\Gamma_{\max}(d-1,1)=\kappa^{d-1}n^{\gamma(3-\beta)(d-1)}$ as defined in \prettyref{eq:Gammadku_upperbound} and $\kappa=\frac{(1+2\delta)^22^{5-\beta}C}{(2^{3-\beta}-1)\overline{w}}$. 
\end{lemma}
\begin{remark}
We provide an intuitive explanation on $\Psi(d,k).$ Analogous to \prettyref{rmk:boundfakepairinP1}, there are two extreme cases in which a vertex $i$ in $P_k$ becomes a common $d$-hop neighbor of $(u,v)$. One case is that $i$ connects to some $(d-1)$-hop neighbor of $u$ and $v$, respectively. Recall that $\Gamma_{\max}(d-1,l)$ is  an upper bound of its $(d-1)$-hop neighbors in $P_l$ by~\prettyref{lmm:Omega}. 
Thus, a vertex $i$ in $P_k$ 
connects to at least one $(d-1)$-hop neighbor of $u$ with probability at most $\sum_{l=1}^{K}\Gamma_{\max}(d-1,l)\frac{\alpha_k\alpha_l}{n\overline{w}}\approx  \Gamma_{\max}(d-1,1)\alpha_{k}n^{\gamma-1}$, where the approximation holds because $l=1$ is the dominating term in the summation. Moreover, there are $\Theta(n\alpha_{k}^{1-\beta})$ vertices in the slice $P_k$. Thus, for a fake pair $(u,v)$, its number of such common $d$-hop neighbors in $P_k$ is about $\Gamma_{\max}^2(d-1,1)n^{2\gamma-1} \alpha_k^{3-\beta}$,
which gives rise to the first term of $\Psi(d,k).$ The other extreme case is that $i$ is a $(d-1)$-hop neighbor of some common neighbor of $(u,v)$. As stated in \prettyref{rmk:boundfakepairinP1}, we can bound from above $\abs{\Gamma_{1}^{\overline{G}_1}(u)\cap \Gamma_{1}^{\overline{G}_2}(v)}$ by $\log n$. Then, the vertex $i$ connects to at least one $(d-2)$-hop neighbor of a given vertex in $\Gamma_{1}^{\overline{G}_1}(u)\cap \Gamma_{1}^{\overline{G}_2}(v)$ with probability at most $\sum_{l=1}^{K}\Gamma_{\max}(d-2,l)\frac{\alpha_k\alpha_l}{n\overline{w}}\approx  \Gamma_{\max}(d-2,1)\alpha_{k}n^{\gamma-1}$. Again, there are $\Theta(n\alpha_{k}^{1-\beta})$ vertices in the slice $P_k$. Thus, the number of such common $d$-hop neighbors in $P_k$ is about $\Gamma_{\max}(d-2,1)\alpha_{k}^{2-\beta}n^{\gamma}\log n\approx 2^{(k-1)(\beta-2)}\Gamma_{\max}(d-1,1)\log n$, which gives rise to the second term of $\Psi(d,k).$
\end{remark}


Before proving~\prettyref{lmm:Gammauvk}, we first show how
to apply~\prettyref{lmm:Gammauvk} to prove~\prettyref{lmm:boundfakepairinP1}.
combining \prettyref{eq:set_inclusion_D}, \prettyref{eq:Ndku}, and \prettyref{eq:Gammauvk} yields that 
$$
\prob{\abs{\Gamma_{d,k}^{\overline{G}_1}(u)\cap \Gamma_{d,k}^{\overline{G}_2}(v)}\le \Psi(d,k)+2N_{\max}(d-1,k)}>1-n^{-4+o(1)}.
$$

Next we set $d=D$ and sum over $k$ for all the slices $P_k$ with weight at most $\frac{15}{s}\log n$, \ie, $\alpha_{k}\le\frac{15}{s}\log n$. 
In particular, we have $k\ge k_0\triangleq \lfloor\log_2(\frac{n^{\gamma}s}{15\log n}) \rfloor$ and
\begin{align*}
&\sum_{k= k_0}^{K}
\Psi(D,k)+2N_{\max}(D-1,k)\\
\leq & \sum_{k=  k_0}^{K}\frac{\kappa^2\Gamma_{\max}^2(D-1,1)n^{\gamma(5-\beta)}}{ 2^{(k-1)(3-\beta)}Cn}+\frac{\Gamma_{\max}(D-1,1)}{2^{(k-1)(2-\beta)}}6\log n+\frac{4\kappa^{D-1}n^{\gamma(3-\beta)(D-1)}}{2^{(k-1)(2-\beta)}}\\
\le & \frac{2^{3-\beta}\kappa^2\Gamma_{\max}^2(D-1,1)n^{\gamma(5-\beta)}}{(2^{3-\beta}-1)  2^{(k_0-1)(3-\beta)}Cn}  
+\frac{2^{\beta-2}}{2^{\beta-2}-1}\frac{\Gamma_{\max}(D-1,1)}{2^{(K-1)(2-\beta)}}6\log n+\frac{2^{\beta-2}}{2^{\beta-2}-1} \frac{4\kappa^{D-1}n^{\gamma(3-\beta)(D-1)}}{2^{(K-1)(2-\beta)}}  \\
\le& \frac{2^{3-\beta}\kappa^{2D}n^{2\gamma((3-\beta)(D-1)+1)}}{(2^{3-\beta}-1) Cn}\left(\frac{15}{s}\log n\right)^{3-\beta}+\frac{2^{\beta-2}}{2^{\beta-2}-1}\kappa^{D-1} n^{(\gamma(3-\beta)(D-2)+1)}(4+6\log n)=\Psi_{\max},
\end{align*}
where $\Psi_{\max}$ is given in \prettyref{eq:Gammauvd}.
Thus, we prove  the desired conclusion \prettyref{eq:Gammauvd}.

Next we present the proof of \prettyref{lmm:Gammauvk}.
\begin{proof}[Proof of \prettyref{lmm:Gammauvk}]
Fix two  distinct vertices $u,v$ in $\overline{P}_1$, we study their common $d$-hop neighborhood  from $d=1$. 
\paragraph{For $d=1$:} For each vertex $i\in \overline{P}_k$, define an indicator variable 
$$
x_i^k=\indc{i\in \Gamma_1^{\overline{G}_1}(u) \cap  \Gamma_1^{\overline{G}_2}(v)}.
$$
In other words, $x_i^k=1$ if $i$ is connected to $u$ in $\overline{G}_1$ and $v$ in $\overline{G}_2$, and $x_i^k=0$ otherwise. Then, we have $\abs{\Gamma_{1,k}^{\overline{G}_1}(u)\cap \Gamma_{1,k}^{\overline{G}_2}(v) }=\sum_{i\in \overline{P}_k}x_i^k$. Since $w_u,w_v\in[(1-2\delta]\alpha_{1},(1+2\delta)\alpha_0]$, it follows that
\begin{align*}
    \prob{x_i^k=1}\leq \left((1+2\delta)^2\frac{\alpha_{k-1}\alpha_0}{n\overline{w}}\right)^2
    \triangleq p_{\max}^k.
\end{align*}

Hence, we have 
$$
\abs{\Gamma_{1,k}^{\overline{G}_1}(u)\cap \Gamma_{1,k}^{\overline{G}_2}(v)}\overset{s.t.}{\leq} \Binom\left(\abs{\overline{P}_k},p_{\max}^k\right).
$$

Recall  $n_k=Cn\alpha^{1-\beta}_{k-1}$ in view of \prettyref{eq:Pknumber} and $ \abs{\overline{P}_k}\leq 2n_k$ in view of \prettyref{eq:numberPkbar}. Hence,
$$
2n_kp_{\max}^k=(1+2\delta)^4\frac{2C\alpha_{k-1}^{3-\beta}n^{2\gamma}}{\overline{w}^2n}.
$$
 Hence, we apply \prettyref{lmm:nkplus1issue} with $\lambda=4\log n$, and get
\begin{align*}
    &\prob{\abs{\Gamma_{1,k}^{\overline{G}_1}(u)\cap \Gamma_{1,k}^{\overline{G}_2}(v)}\geq \frac{4(1+2\delta)^4C\alpha_{k-1}^{3-\beta}n^{2\gamma}}{\overline{w}^2n}+\frac{16}{3}\log n}\le n^{-4}.
\end{align*}
Since $\Gamma_{\max}(0,1)=1$, we have $\Psi(1,k)=\frac{\kappa^2\alpha_{k-1}^{3-\beta}n^{2\gamma}}{Cn}+6\log n$. Thus, \prettyref{eq:Gammauvk} holds for $d=1$.

\paragraph{For $2\le d\le D$:} We first count the $d$-hop neighbors conditional on the $(d-1)$-hop neighborhood of $u$ and $v$. We use $\Omega_{d}^*$ to denote the event that $\Omega_{d-1}$ with $\overline{G}=\overline{G}_1,\overline{G}_2$ hold, and for all $k=1, \ldots, K$,
\begin{align*}
   \abs{\Gamma_{d,k}^{\overline{G}_1}(u)\cap \Gamma_{d,k}^{\overline{G}_2}(v)}\le\Psi(d,k),
\end{align*}
with $\Psi(d,k)$ defined in \prettyref{lmm:Gammauvk}.

Conditioning on $\Omega_{d-1}^*$, note that there are two possible cases under which each true pair $(i,i)$ becomes a common $d$-hop neighbor of $(u,v)$. One case is that $i$ connects to some common $(d-1)$-hop neighbors of $(u,v)$ in both $\overline{G}_1$ and $\overline{G}_2$. The other case is that $i$ connects to different $(d-1)$-hop neighbors of $(u,v)$ in $\overline{G}_1$ and $\overline{G}_2$, respectively. 

For each vertex $i\in \overline{P}_k\setminus N_{D-1}(u,v)$, define two  indicator variables 
\begin{align*}
 y_i^k=&\indc{i\in \Gamma_d^{\overline{G}_1}(u),i\in \Gamma_d^{\overline{G}_2}(v)},\\
 z_i^k=&\indc{\exists j\in \Gamma_{d-1}^{\overline{G}_1}(u)\cap \Gamma_{d-1}^{\overline{G}_2}(v):\; i\in \Gamma_1^{\overline{G}_1}(j)}. 
\end{align*}
In other words, $y_i^k=1$ if $i$ is a $d$-hop neighbor of $u$ in $\overline{G}_1$ and $v$ in $\overline{G}_2$, and $y_i^k=0$ otherwise. Similarly, $z_i^k=1$ if $i$ is connected to at least one common $(d-1)$-hop neighbor of $(u,v)$ in both $\overline{G}_1$ and $\overline{G}_2$, and $z_i^k=0$ otherwise. Note that $z_i^k=1$ includes the case that  $i$ connects to some common $(d-1)$-hop neighbors of $(u,v)$ in both $\overline{G}_1$ and $\overline{G}_2$.

We first bound $\prob{z_i^k=1|\Omega_{d-1}^*}$ from above by
\begin{align*}
\prob{z^k_i=1\mid\Omega_{d-1}^*}\overset{(a)}{\le}&\sum_{l=1}^{K}\prob{\exists j\in \Gamma_{d-1}^{\overline{G}_1}(u)\cap \Gamma_{d-1}^{\overline{G}_2}(v):  i\in \Gamma_1^{\overline{G}_1}(j)\mid\Omega_{d-1}^*}\\
\overset{(b)}{\leq}& (1+2\delta)^2\sum_{l=1}^{K} \Psi(d-1,l)\frac{\alpha_{k-1}\alpha_{l-1}}{n\overline{w}}\\
\le&\left(\frac{\kappa^2\Gamma_{\max}^2(d-2,1)n^{\gamma(7-\beta)}}{ 2^{k-1}Cn^2\overline{w}}+\frac{6\Gamma_{\max}(d-2,1)n^{2\gamma}\log n}{2^{k-1}n\overline{w}}\right)\sum_{l=1}^{K}\frac{(1+2\delta)^2}{2^{(l-1)(3-\beta)}}\\
\overset{}{\leq} &  \frac{\kappa^{2d-1}n^{2\gamma(3-\beta)(d-2)}n^{\gamma(7-\beta)}}{ 2^{k+1}C^2n^2}+\frac{6\kappa^{d-1}n^{\gamma((3-\beta)(d-2)+2)}\log n}{2^{k+1}Cn}=\nu_1,
\end{align*}
where $(a)$ holds due to the union bound; $(b)$ follows from the union bound and event $\Omega_{d-1}^*$. 

Then, the event $\{y_i^k=1\}\setminus \{z_i^k=1\}$ denotes the event that $i$ connects to some vertex in $\Gamma_{d-1,k}^{\overline{G}_1}(u)\setminus \Gamma_{d-1,k}^{\overline{G}_2}(v)$  and connects to some vertex in $\Gamma_{d-1,k}^{\overline{G}_2}(v)$ independently. Thus, $\prob{\{y_i^k=1\}\setminus \{z_i^k=1\}\mid\Omega_{d-1}^*}$ can be bounded by
\begin{align*}
&\prob{\{y_i^k=1\}\setminus \{z_i^k=1\}\mid\Omega_{d-1}^*}\\
\le & \prob{\exists j\in \Gamma_{d-1,k}^{\overline{G}_1}(u)\setminus \Gamma_{d-1,k}^{\overline{G}_2}(v):i\in \Gamma_1^{\overline{G}_1}(j)\mid\Omega_{d-1}^* }\prob{\exists j\in  \Gamma_{d-1}^{\overline{G}_2}(v):  i\in \Gamma_1^{\overline{G}_2}(j)\mid\Omega_{d-1}^*}\\
\leq &\prob{i\in\Gamma_{d}^{\overline{G}_1}(v)\mid\Omega_{d-1}^*}\prob{i\in\Gamma_{d}^{\overline{G}_2}(v)\mid\Omega_{d-1}^*}\\
\overset{(a)}{\le}&\left(\frac{\kappa^{d}n^{\gamma((3-\beta)(d-1)+2)}}{2^{k+1}Cn}\right)^2\\
\le&
\frac{\kappa^{2d}n^{2\gamma(3-\beta)(d-2)}}{2^{2(k+1)}C^2n^2}n^{2\gamma(5-\beta)}=\nu_2,
\end{align*}
where $(a)$ follows from a similar proof of \prettyref{eq:pmaxkd}.

When we compare the first term of $\nu_1$ and $\nu_2$, we have
$$
\frac{n^{2\gamma(3-\beta)(d-2)}}{n^2}n^{\gamma(7-\beta)}\ll \frac{n^{2\gamma(3-\beta)(d-2)}}{n^2}n^{2\gamma(5-\beta)},
$$
where the inequality follows from $\frac{n^{\gamma(7-\beta)}}{n^{2\gamma(5-\beta)}}=n^{\gamma(\beta-3)}=o( 1)$.

Thus, we have
\begin{align*}
\prob{y_i^k=1\mid \Omega_{d-1}^* }
& \leq \nu_1+\nu_2\le 2\nu_2+\frac{4\kappa^dn^{\gamma((3-\beta)(d-2)+2)}\log n}{2^{k}Cn} \\
& \le \frac{\kappa^{2d}n^{2\gamma(3-\beta)(d-1)}n^{4\gamma}}{ 3\cdot2^{2(k-1)}C^2n^2}+\frac{4\kappa^dn^{\gamma((3-\beta)(d-2)+2)}\log n}{2^{k}Cn}\triangleq\mu_k.
\end{align*}

Thus, conditional on $\Omega_{d-1}^*$, we have 
$$\abs{\Gamma_{d,k}^{\overline{G}_1}(u)\cap \Gamma_{d,k}^{\overline{G}_2}(v)}\overset{s.t.}{\leq}\Binom \left(\abs{\overline{P}_k},\mu_k\right).
$$ 
Recall  $n_k=Cn\alpha^{1-\beta}_{k-1}$ in view of \prettyref{eq:Pknumber} and $ \abs{\overline{P}_k}\leq 2n_k$ in view of \prettyref{eq:numberPkbar}. Therefore, for sufficiently large $n$,
$$
2n_k\mu_k =\frac{2\kappa^{2d}n^{2\gamma(3-\beta)(d-1)}n^{\gamma(5-\beta)}}{ 3\cdot2^{(k-1)(3-\beta)}Cn}+\frac{4\kappa^{d-1}n^{\gamma(3-\beta)(d-1)}\log n}{2^{(k-1)(2-\beta)}}\le \frac{2}{3}\Psi(d,k).
$$
We then apply Chernoff Bound with $\eta=\frac{1}{2}$ and get
\begin{align*}
  \prob{\abs{\Gamma_{d,k}^{\overline{G}_1}(u)\cap \Gamma_{d,k}^{\overline{G}_2}(v)}\ge \Psi_{\max}(d,k)\mid \Omega_{d-1}^*}\leq n^{-4}.
\end{align*}

\paragraph{Induction:} Finally, we prove \prettyref{eq:Gammauvk} by induction.

For $d=1$, we have proved that \prettyref{eq:Gammauvk} holds.

Suppose  \prettyref{eq:Gammauvk}  holds for $d-1$, then taking the union bound yields that
\begin{align*}
    \prob{\Omega_{d-1}^{*c}}\le& 2\cdot\prob{\Omega_{d-1}^c}+\prob{\abs{\Gamma_{d-1,k}^{\overline{G}_1}(u)\cap \Gamma_{d-1,k}^{\overline{G}_2}(v)}\ge\Psi_{\max}(d-1,k)}\\
    \leq& 2(4^{d-1}-1)n^{-4}+\frac{2\cdot 4^{d-1}}{3}n^{-4}= \left(\frac{2\cdot4^{d}}{3}-1\right) \cdot n^{-4}.
\end{align*}

Thus,  we have
\begin{align*}
   & \prob{\abs{\Gamma_{d,k}^{\overline{G}_1}(u)\cap \Gamma_{d,k}^{\overline{G}_2}(v)}\geq \Psi_{\max}(d,k)}\\
   \leq &\prob{\Omega_{d-1}^{*c}}+ \prob{\abs{\Gamma_{d,k}^{\overline{G}_1}(u)\cap \Gamma_{d,k}^{\overline{G}_2}(v)}\ge \Psi_{\max}(d,k)\mid \Omega_{d-1}^*}\\
   \leq& \left(\frac{2\cdot4^{d}}{3}-1\right) \cdot n^{-4}+n^{-4}
   \leq  \frac{2\cdot4^{d}}{3}  \cdot n^{-4} .
\end{align*}
\end{proof}

\subsubsection{Proof of \prettyref{lmm:proofP1}}\label{sec:pf-lmm3}
The main idea of the proof is to bound 
the number of $D$-hop witnesses for both true pairs and fake pairs in the first slice,
using the bounds to the number of the $D$-hop neighbors 
established in \prettyref{lmm:boundtruedhopkslice} and \prettyref{lmm:boundfakepairinP1}.

Recall that in Algorithm \prettyref{alg:D-hop-power-law}, 
we select the set $\hat{\calS}$ of low-degree seeds. 
Let $\hat{S}=\{i: (i,i) \in \hat{\calS}\}$. 
To circumvent the dependency between  $\hat{S}$ and the graphs $G_1$ and $G_2$,
we will introduce $ \underline{S}$ and $\overline{S}$
such that they are independent from graphs
and $ \underline{S} \subset \hat{S} \subset \overline{S}$
with high probability. To this end, we define an event $\calE$ such that 
$$
\{i: w_i \le c\} \subset \{ i: |\Gamma_1^{G_1} (i) | \le 5 \log  n , |\Gamma_1^{G_2} (i) | \le 5 \log n \} \subset \{i: w_i \le \frac{15}{s} \log n\}.
$$

For any $i$ with $w_i\le c$, $\expect{|\Gamma_1^{G_1} (i) |}=cs$. Thus, applying \prettyref{lmm:nkplus1issue} with $\lambda=3\log n$ yields
$$
\prob{|\Gamma_1^{G_1} (i) | \ge 5\log n}\le \prob{|\Gamma_1^{G_i} (i) | \ge 2cs+4\log n}\le n^{-3}.
$$
Taking a union bound over $i$ gives $\prob{\{i: w_i \le c\} \subset \{ i: |\Gamma_1^{G_1} (i) | \le 5 \log  n , |\Gamma_1^{G_2} (i) | \le 5 \log n \}}\ge 1-n^{-2+o(1)}$.

For any $i$ with $w_i> \frac{15}{s}\log n$,  $\expect{|\Gamma_1^{G_1} (i) |}=15\log n$.  we apply Chernoff Bound in \prettyref{thm:chernoffbound} with 
$\eta=2/3$ and have
$$
\prob{|\Gamma_1^{G_1} (i) | \le 5\log n}\le \prob{|\Gamma_1^{G_i} (i) | \le \left(1-\frac{2}{3}\right){15}\log n}\le n^{-3}.
$$
Thus, we have
\begin{align*}
&\prob{ \{ i: |\Gamma_1^{G_1} (i) | \le 5 \log  n , |\Gamma_1^{G_2} (i) | \le 5 \log n \} \subset \{i: w_i \le \frac{15}{s} \log n\}}\\
= &\prob{ \{i: w_i > \frac{15}{s} \log n\}\subset\{ i: |\Gamma_1^{G_1} (i) | > 5 \log  n , |\Gamma_1^{G_2} (i) | >5 \log n \} }=1-n^{-2+o(1)}.
\end{align*}

Thus, $\prob{\calE} \ge 1- n^{-2+o(1)}.$ On event $\calE$, we have
$$
 \underline{S} \triangleq\{i:w_i\le c\}\cap S \subset \hat{S} \subset \{i:w_i\le \frac{15}{s}\log n \}\cap S \triangleq \overline{S},
$$
where $S=\{i: (i,i) \in \calS\}$ denotes the set of vertices selected as the initial seed
set $\calS$. Note that crucially the initial seeds in $\calS$ are selected among all true pairs with probability $\theta$, independently from everything else. 
Thus $\underline{S}$ and $\overline{S}$ are independent from graphs. 
As a consequence, to bound from below (resp.\ above) the number of $D$-hop witnesses 
for the true (resp.\ fake) pair, it suffices to consider their common $D$-hop neighbors in 
$\underline{S}$ (resp.\ $\overline{S}$).

More specifically, let us first consider the true pairs. 
Fix any vertex $u\in P_1$. Let $\Lambda(u)=\Gamma_D^{\underline{G}_1}(u)\cap\Gamma_D^{\underline{G}_2}(u)\setminus \left(N_{D-1}^{\overline{G}_1}(u)\cap N_{D-1}^{\overline{G}_2}(u)\right)$.
Define event 
\begin{align*}
    \calA_u=\left\{\abs{\Lambda(u)\cap \underline{S} }> \frac{3}{5}\Gamma_{\min}\theta\right\}, \text{ where } \Gamma_{\min}=\frac{1}{2}\left(\frac{C\cdot s^2}{12\cdot\overline{w}}\right)^{D}n^{\gamma((3-\beta)(D-1)+1)}.
\end{align*}

Note that due to assumption \prettyref{eq:locallytreelikecondition} and $n^{\gamma(3-\beta)}\gg \log n$ for sufficiently large $n$,
$N_{\max} \le 
\frac{1}{10} \Gamma_{\min}$. 
Hence it follows from \prettyref{lmm:boundtruedhopkslice} that
$$
\prob{\abs{ \Lambda(u)\cap \{i:w_i\le c}<\frac{4}{5}\Gamma_{\min}}\leq n^{-4+o(1)}.
$$

Because the seeds $\calS$ are selected among all true pairs with probability $\theta$, independently from everything else, we have
$$
\abs{\Lambda(u)\cap \underline{S}}\sim\Binom\left(\abs{\Lambda(u)\cap\{i:w_i\le c\} },\theta\right).
$$
Then, we apply Chernoff Bound in \prettyref{thm:chernoffbound} with $\eta=\frac{1}{4}$ and get
\begin{align*}
   \prob{\calA_u^c}\leq& \prob{\abs{ \Lambda(u)\cap \{i:w_i\le c\}}<\frac{4}{5}\Gamma_{\min}}+\prob{\calA_u^c \,\bigg|\, \abs{\Lambda(u) \cap \{i:w_i\le c\}}\ge \frac{4}{5}\Gamma_{\min}}\\
\overset{}{\le} &n^{-4+o(1)}+\prob{\Binom\left(\Gamma_{\min},\theta\right)\le \frac{3}{5}\Gamma_{\min}\theta}\\
\le&n^{-4+o(1)}+\exp\left(-\frac{1}{40}\Gamma_{\min}\theta\right)
\overset{(a)}{\le}  n^{-4+o(1)},
\end{align*}
where  $(a)$ holds due to assumption \prettyref{eq:theta}.
Let $\calA=\cap_{u \in P_1} \calA_u$. 
It follows from the union bound that
$\prob{\calA} \le n^{-3+o(1)}.$


We next consider the fake pairs. Fix any two distinct vertices $u,v\in \overline{P}_1$.
Define an event 
\begin{align*}
  \calB_{uv}=\left\{\abs{N_D^{\overline{G}_1}(u)\cap N_D^{\overline{G}_2}(v)\cap \overline{S}}\leq \frac{1}{2}\Gamma_{\min}\theta\right\}.
\end{align*}

Note that due to the assumption \prettyref{eq:locallytreelikecondition} and $n^{\gamma(3-\beta)}\gg \log n$ for sufficiently large $n$,
$$
\Psi_{\max}\leq \frac{\kappa^{2D}}{\left(\frac{Cs^2}{12\cdot \overline{w}}\right)^D}\left(\frac{2^{3-\beta}n^{\gamma((3-\beta)(D-1)+1)}}{(2^{3-\beta}-1)Cn}\left(\frac{15}{s}\log n\right)^{3-\beta}+\frac{2^{\beta-2}(4+6\log n)}{(2^{\beta-2}-1)n^{\gamma(3-\beta)}}\right)\Gamma_{\min}\leq \frac{1}{4}\Gamma_{\min}.
$$
Hence, it follows from \prettyref{lmm:boundfakepairinP1} that 
$$
\prob{\abs{N_D^{\overline{G}_1}(u)\cap N_D^{\overline{G}_2}(v)\cap \{i:w_i\le \frac{15}{s}\log n\}}>\frac{1}{4}\Gamma_{\min}}\leq n^{-4+o(1)}.
$$
Since the seeds $\calS$ are selected among all true pairs with probability $\theta$ independently, we have
$$
\abs{\Gamma_D^{\overline{G}_1}(u)\cap\Gamma_D^{\overline{G}_2}(v)\cap \overline{S}}\sim\Binom\left(\abs{N_D^{\overline{G}_1}(u)\cap N_D^{\overline{G}_2}(v)\cap \{i:w_i\le \frac{15}{s}\log n\}},\theta\right).
$$

Then, we apply Chernoff Bound in \prettyref{thm:chernoffbound} with $\eta=1$ and get
\begin{align*}
   \prob{\calB_{uv}^c}\leq& \prob{\abs{ N_D^{\overline{G}_1}(u)\cap N_D^{\overline{G}_2}(v)\cap \{i:w_i\le \frac{15}{s}\log n\}}>\frac{1}{4}\Gamma_{\min}}\\&+\prob{\calE_{uv}^c \,\bigg|\, \abs{ N_D^{\overline{G}_1}(u)\cap N_D^{\overline{G}_2}(u)\cap \{i:w_i\le \frac{15}{s}\log n\}}\le \frac{1}{4}\Gamma_{\min}}\\
\overset{}{\le} &n^{-4+o(1)}+\prob{\Binom\left(\frac{1}{4}\Gamma_{\min},\theta\right)\le \frac{1}{2}\Gamma_{\min}\theta}\\
\le&n^{-4+o(1)}+\exp\left(-\frac{1}{12}\Gamma_{\min}\theta\right)
\overset{(a)}{\le}  n^{-4+o(1)},
\end{align*}
where  $(a)$ holds due to assumption \prettyref{eq:theta}. 
Let $\calB=\cap_{u,v \in \overline{P}_1: u\neq v} \calB_{uv}$.
It follows from the union bound that 
$\prob{\calB^c} \le n^{-2+o(1)}.$

Finally, we define event $\calC$ such that 
$$
\underline{G}_j\subset \hat{G}_j\subset \overline{G}_j, \quad \forall j=1,2  \quad \text{ and } \quad P_1 \subset \hat{P}_1 \subset \overline{P}_1. 
$$
It follows from \prettyref{lmm:uinPk} that $\prob{\calC}\ge 1-n^{-4+o(1)}.$
Taking the union bound, we have 
$$
\prob{\calA \cap \calB \cap \calC \cap \calE }
\ge 1-n^{-3+o(1)} - n^{-2+o(1)} -2 n^{-4+o(1)} \ge 1-n^{-2+o(1)}.
$$
It remains to verify that on the event 
$\calA \cap \calB \cap \calC \cap \calE$, 
$\calR_1$ contains 
all true pairs in $Q_1$ and
no fake pairs in $\hat{Q}_1$.

Recall that we uses seeds in $\hat{\calS}$ and count the $D$-hop witnesses in $\hat{G}_1$ and $\hat{G}_2$ for all candidate vertex pairs in $\hat{Q}_1$ in Step 4 of Algorithm \ref{alg:D-hop-power-law}.  
On event $\calA \cap \calC \cap \calE$,  $\Lambda(u)\subset\Gamma^{\hat{G}_1}_D(u)\cap \Gamma^{\hat{G}_2}_D(u)$ and
the minimum number of $D$-hop witnesses among all true pairs $(u,u)$ in $Q_1$
is lower bounded by $\frac{3}{5}\Gamma_{\min}\theta$.
On event $\calB \cap \calC \cap \calE$, $\Gamma^{\hat{G}_1}_D(u)\cap \Gamma^{\hat{G}_2}_D(v)\subset N^{\overline{G}_1}_D(u)\cap N^{\overline{G}_2}_D(v)$
the maximum number of $D$-hop witnesses among all fake pairs $(u,v)$ in $\hat{Q}_1$
is upper bounded by 
$\frac{1}{2}\Gamma_{\min}\theta
$. Thus, GMWM with threshold $\tau_1=\frac{1}{2}\Gamma_{\min}\theta$ outputs 
$\calR_1$, which contains 
all true pairs in $Q_1$ and
no fake pairs in $\hat{Q}_1$.




\subsubsection{Proof of \prettyref{lmm:lowerbound-Pk-true}}\label{sec:pf-lmm4}
 Fix a vertex $u \in P_k$. 
 For each vertex $i\in P_{k-1}$,  let $x_i$ be a binary random variable such that $x_i=1$ if $i$  connects to $u$ both in $G_1$ and $G_2$, and $x_i=0$ otherwise. Then, 
 $\abs{\Gamma_{1}^{G_1}(u)\cap \Gamma_{1}^{G_2}(u)\cap P_{k-1}}=\sum_{i \in P_{k-1}} x_i$
 and $x_i$'s are independent. Moreover, we have
 \begin{align*}
    \prob{x_i=1}\geq \frac{\alpha_{k}\alpha_{k-1}}{n\overline{w}}s^2.
\end{align*}
Therefore, 
applying Chernoff Bound in \prettyref{thm:chernoffbound} with $\eta=\frac{1}{2}$ yields that
\begin{align*}
&\prob{\abs{\Gamma_{1}^{{G}_1}(u)\cap \Gamma_{1}^{{G}_2}(u)\cap P_{k-1}}\leq \frac{C\alpha_{k-1}^{3-\beta}s^2}{16\overline{w}}}
\le \prob{\Binom\left(n_{k-1},\frac{\alpha_{k}\alpha_{k-1}}{n\overline{w}}s^2\right)
\le\frac{C\alpha_{k-1}^{3-\beta}s^2}{2^{\beta+1}\overline{w}}} \le n^{-3},
\end{align*}
where the last inequality holds because
$
n_{k-1}\frac{\alpha_{k}\alpha_{k-1}}{n\overline{w}}s^2= \frac{C\alpha_{k-1}^{3-\beta}s^2}{2^{\beta}\overline{w}} \ge 24 \log n
$
in view of $ \left( \alpha_{k^*} \right)^{3-\beta} \ge \frac{192 \overline{w} \log n}{Cs^2}$.

\subsubsection{Proof of \prettyref{lmm:upperbound-Pk-fake}}\label{sec:pf-lmm5}
Fix a pair of two distinct vertices $u, v \in \overline{P}_k$. 
For each vertex $i\in \overline{P}_{k-1}$,  let $x_i$ be a binary random variable such that $x_i=1$ if $i$ is connected to $u$ in $G_1$ and $v$ in $G_2$, and $x_i=0$ otherwise. Since the event that $i$ is connected to $u$ is independent of the event that $i$ is connected to $v$, we have 
\begin{align*}
    \prob{x_i=1}\leq \left((1+2\delta)^2\frac{\alpha_{k-1}\alpha_{k-2}}{n\overline{w}}s\right)^2
    =\frac{4(1+2\delta)^4 \alpha_{k-1}^4s^2 }{ n^2 \overline{w}^2}
    \triangleq p_{\max}.
\end{align*}
Moreover, $x_i$'s are independent. Therefore, 
$\abs{\Gamma_{1}^{G_1}(u)\cap \Gamma_{1}^{G_2}(v)\cap \overline{P}_{k-1}}\overset{s.t.}{\leq} \Binom \left(\abs{\overline{P}_{k-1}},p_{\max}\right)$. 
Recall  $n_{k-1}=Cn\alpha^{1-\beta}_{k-2}$ in view of \prettyref{eq:Pknumber} and $\abs{\overline{P}_{k-1}}\leq 2n_{k-1}$ in view of \prettyref{eq:numberPkbar}. Thus, we 
apply \prettyref{lmm:nkplus1issue} with $\lambda=4\log n$, and get
\begin{align*}
    &\prob{\abs{\Gamma_{1}^{G_1}(u)\cap \Gamma_{1}^{G_2}(v)\cap \overline{P}_{k-1}}\geq \frac{8(1+2\delta)^4C\alpha_{k-1}^{5-\beta}s^2}{\overline{w}^2n}+\frac{16}{3}\log n}\le n^{-4}.
\end{align*}

\subsubsection{Proof of \prettyref{lmm:proofPk}}\label{sec:pf-lmm6}

The proof is divided into two parts.
The first part is to identify a set of ``good'' events whose intersection holds
with high probability. The second part provides a deterministic argument,
showing that on the intersection of these
good events, the $1$-hop algorithm successfully matches slice $k$ for all $2 \le k \le k^*.$

First, we identify a good event under which the number of common  $1$-hop neighbors of a
true pair  is large. More precisely, 
for any vertex $u\in P_k$, define event
\begin{align*}
    \mathcal{A}_k(u)=\left\{\abs{\Gamma_{1}^{G_1}(u)\cap \Gamma_{1}^{G_2}(u)\cap P_{k-1}}\geq \xi_{k} \right\}, \quad \text{ where }  \; \xi_{k} \triangleq \frac{C\alpha_{k-1}^{3-\beta}s^2}{16\overline{w}},
\end{align*}
and $\calA=\cap_{2 \le k \le k^*}\cap_{u \in P_k} \calA_k(u)$.
By \prettyref{lmm:lowerbound-Pk-true} and union bound, we have 
$\prob{\calA^c} \le n^{-2+o(1)}.$

Second, we determine a good event under which the number of common $1$-hop neighbors of 
a fake pair is small. More formally, for any pair of distinct vertices $u, v\in \overline{P}_k$,
define event 
\begin{align*}
   \mathcal{B}_{k}(u,v)=\left\{\abs{\Gamma_{1}^{G_1}(u)\cap \Gamma_{1}^{G_2}(v)\cap \overline{P}_{k-1}}\leq \zeta_{k} \right\},
   \quad \text{ where }  \; \zeta_{k} \triangleq \frac{8(1+2\delta)^4C\alpha_{k-1}^{5-\beta}s^2}{\overline{w}^2n}+\frac{16}{3}\log n,
\end{align*}
and $\calB=\cap_{2 \le k \le k^*} \cap_{u,v \in \overline{P}_k: u\neq v} \calB_{k}(u,v)$.
By \prettyref{lmm:upperbound-Pk-fake} and union bound, 
we have $\prob{\calB^c} \le n^{-3+o(1)}.$

Third, we define an event 
$
\calC =\cap_{2 \le k \le k^*} \left\{ Q_k\subset\hat{Q}_k \subset \overline{Q}_k \right\}.
$
By \prettyref{lmm:uinPk} and union bound, we have $\prob{\calC^c}\le n^{-4+o(1)}.$

Finally, we let $\calF$ denote the event that the first slice is successfully matched, 
\ie, $\calR_1$ contains all true pairs in $Q_1$ and no fake pairs in $\hat{Q}_1$. 
By~\prettyref{lmm:proofP1}, $\prob{\calF^c} \le n^{-1.5+o(1)}.$

Combining the above, it follows that 
$$
\prob{ \calA \cap \calB \cap \calC \cap \calF } \ge 1-n^{-2+o(1)} -n^{-3+o(1)} - n^{-4+o(1)} -n^{-1.5+o(1)}
\ge 1-n^{-1.5+o(1)}.
$$
It remains to verify on the event $\calA \cap \calB \cap \calC \cap F$, 
$\calR_k$ contains
all true pairs in ${Q}_k$ and no fake pairs in $\hat{Q}_k$ for all $1\leq k\leq k^*$.
We prove this by induction. The base case with $k=1$ follows from the definition of $\calF.$
Assume the induction hypothesis holds for the slice $k-1$, we aim to show it continues to hold for $k.$

Recall that when matching the slice $\hat{Q}_k$, we use $\calR_{k-1}$ as the set of seeds. 
Since the induction hypothesis is true for slice $k-1$, it follows that 
$\calR_{k-1}$ contains all the true pairs in $Q_{k-1}.$ 
Thus, 
the minimum number of $1$-hop witnesses among all true pairs $(u,u)$ in $Q_k$
is lower bounded by $\xi_{k}.$
Moreover, since $\calR_{k-1}$ contains no fake pairs in $\hat{Q}_{k-1}$
and on event $\calC$, $\hat{Q}_{k-1} \subset \overline{Q}_k$, 
it follows that 
$\calR_{k-1}$ is contained by all the true pairs in $\overline{Q}_{k-1}$.
Also, the set of fake pairs  in $\hat{Q}_k$ is contained by
the set of fake pairs in $\overline{Q}_k$. 
Thus, 
the maximum number of $1$-hop witnesses among all fake pairs $(u,v)$ in $\hat{Q}_k$
is upper bounded by 
$\zeta_{k}.
$


Note that
$$
\xi_k\overset{(a)}{\ge} \tau_2(k) \text{ and } \frac{\zeta_{k}}{\tau_2(k)}\overset{(b)}{
\leq} \frac{128(1+2\delta)^4n^{2\gamma}}{\overline{w}n}+\frac{4}{9}\overset{(c)}{<}1,
$$
where $(a)$ holds by definition of $\tau_2(k)$ in \prettyref{eq:def_tau_2}; $(b)$ follows from $n^{\gamma}\ge\alpha_k\ge\alpha_{k^*}\geq\left(\frac{192\overline{w}\log n}{Cs^2}\right)^{\frac{1}{3-\beta}}$ for $2\le k\le k^*$; $(c)$ holds as $n$ is sufficiently large in view of $n^{2\gamma}=o(n)$
and $\overline{w}=\Theta(1).$
Thus, $\calR_k$ contains 
all true pairs in $Q_k$ and
no fake pairs in $\hat{Q}_k$,
completing the induction.

\subsubsection{Proof of \prettyref{lmm:proofPGMfake}}\label{sec:pf-lmm7}
Fix any two distinct vertices $u, v \in \overline{P}_{\ge k^*+1}$. 
Then $w_u, w_v \le (1+2\delta) \alpha_{k^*}$. 
For each vertex $i\in \overline{P}_{\ge k^*}$,  let $x_i$ be a binary random variable such that $x_i=1$ if $i$  connects to  $u$ in $G_1$ and $v$ in $G_2$, and $x_i=0$ otherwise. Since the event that $i$ connects to $u$ is independent of the event that $i$ connects to $v$, we have 
\begin{align*}
    \prob{x_i=1}\leq \left((1+2\delta)^2\frac{\alpha_{k^*}\alpha_{k^*-1}}{n\overline{w}}s\right)^2
    =(1+2\delta)^4\frac{4 \alpha_{k^*}^4s^2 }{n^2\overline{w}^2}
    \triangleq p_{\max}.
\end{align*}
Moreover, $x_i$'s are independent. Therefore,
$$
\abs{\Gamma_{1}^{G_1}(u)\cap \Gamma_{1}^{G_2}(v)\cap \overline{P}_{\ge k^*} }\overset{s.t.}{\leq}\Binom\left(\abs{\overline{P}_{\ge k^*}},p_{\max}\right)\overset{s.t.}{\leq}\Binom\left(n,p_{\max}\right).
$$
Thus, we get
\begin{align*}
    \prob{\abs{\Gamma_{1}^{G_1}(u)\cap \Gamma_{1}^{G_2}(v)\cap \overline{P}_{\ge k^*}}\geq 3}
    {\leq}&\prob{\Binom\left(n,p_{\max}\right)\geq 3}\\
    \overset{(a)}{\le} &n^3p_{\max}^3\\
    \leq&\frac{64(1+2\delta)^{12}C^3\alpha_{k^*}^{12}s^6}{n^3\overline{w}^6} \leq n^{-3+o(1)}.
\end{align*}
where $(a)$ follows from the union bound.

\subsubsection{Proof of \prettyref{lmm:proofPGMtrue}}\label{sec:pf-lmm8}

We first bound $\abs{S_h}$ by conditioning on $S_{h-1}.$
For any $u\in P_{k^*+h}$, let $x_i$ be a binary random variable such that $x_i=1$ if $i\in S_{h-1}$  connects to $u$, and $x_i=0$ otherwise. Since $S_{h-1}$ is only determined by the vertex weight and the edges connecting to previous $S_l$, $l<h-1$, the event that $i$ and $u$ is connected  is independent across $i$ conditional on $S_{h-1}$. It follows that 
  \begin{align*}
     \prob{x_i=1\mid S_{h-1} }
     \geq \frac{\alpha_{k^*+h}\alpha_{k^*+h-1}}{n\overline{w}}s^2.
 \end{align*}
 
Thus, we have $\abs{\Gamma_{1}^{G_1}(u)\cap\Gamma_1^{G_2}(u)\cap S_{h-1}}\overset{s.t.}{\geq}\Binom\left(\abs{S_{h-1}},\frac{\alpha_{k^*+h}\alpha_{k^*+h-1}}{n\overline{w}}s^2\right)$ conditional on $S_{h-1}$.  Applying Chernoff Bound in \prettyref{thm:chernoffbound} yields that 
\begin{align*}
&\prob{\abs{\Gamma_{1}^{G_1}(u)\cap \Gamma_{1}^{G_2}(u)\cap  S_{h-1}}< 3 \,\bigg|\, \abs{S_{h-1}}\geq \frac{1}{2}n_{k^*+h-1}}\\
    \leq &\prob{\Binom\left(\frac{1}{2}n_{k^*+h-1},\frac{\alpha_{k^*+h}\alpha_{k^*+h-1}}{n\overline{w}}s^2\right)\leq (1-\eta) \mu}\\
    \leq &\exp \left(- \frac{\eta^2}{2} \mu \right)\triangleq p_{h} \le \frac{1}{2\sqrt{2}},
\end{align*}
where $\mu=\frac{1}{2}n_{k^*+h-1}\frac{\alpha_{k^*+h}\alpha_{k^*+h-1}}{n\overline{w}}s^2= \frac{C\alpha_{k^*+h-2}^{3-\beta}s^2}{16\overline{w}}\geq 12 \ln 2$
due to $\alpha_{k^*+h}\geq\left(\frac{192\overline{w}\ln 2}{Cs^2}\right)^{1/(3-\beta)}$ and 
$\eta =\frac{\mu-3}{\mu} \ge \frac{1}{2}$.


Then, the above result implies that: $\expect{\abs{S_h}\mid\abs{S_{h-1}}\geq \frac{1}{2}n_{k^*+h-1}}\geq (1-p_{h}) n_{k^*+h}$. Note that the event $u\in S_h$ only depends on the vertex weight and the edge set  $E_u \triangleq \{(u,i):i\in S_{h-1}\}$ .  Because $E_u$'s are disjoint, the event $u\in S_h$ is independent across $u\in P_{k^*+h}$. Thus, we apply Chernoff Bound in \prettyref{thm:chernoffbound}
with $\eta=\frac{1-2p_h}{2(1-p_h)}$ and have
\begin{align*}
     \prob{\abs{S_h}< \frac{1}{2}n_{k^*+h} \,\bigg|\, \abs{S_{h-1}}\geq \frac{1}{2}n_{k^*+h-1}}
    {\leq}&\prob{\Binom\left(n_{k^*+h},1-p_{h}\right)< \frac{1}{2}n_{k^*+h}}\\
    \leq &\exp\left(-\frac{(1-2p_{h})^2n_{k^*+h}}{8(1-p_h)}\right)
    \leq n^{-3},
\end{align*}
where the last inequality holds due to $n_{k^*+h}\ge n_{k^*}\ge Cn \left(\frac{192\overline{w}\log n}{Cs^2}\right)^{\frac{1-\beta}{3-\beta}}\ge 128 \log n$ 
due to the choice of $k^*$ in~\prettyref{eq:choice_k_star} and sufficiently large $n$.

Finally, we prove by induction that $\prob{\abs{S_h}< \frac{1}{2}n_{k^*+h}}\le h\cdot n^{-3}$.

For $h=0$, it is true by definition.

For $h\ge 1$,  if $\prob{\abs{S_{h-1}}\geq \frac{1}{2}n_{k^*+h-1}}\geq 1-(h-1)\cdot n^{-3}$, then 
\begin{align*}
      \prob{\abs{S_h}< \frac{1}{2}n_{k^*+h}}\leq&\prob{\abs{S_h}< \frac{1}{2}n_{k^*+h}\mid\abs{S_{h-1}}\geq \frac{1}{2}n_{k^*+h-1}}+ \prob{\abs{S_{h-1}}< \frac{1}{2}n_{k^*+h-1}}\\
      \leq&n^{-3}+(h-1)\cdot n^{-3}=h\cdot n^{-3}.
\end{align*}

\subsubsection{Proof of \prettyref{lmm:proofPGM}}
First, for any two distinct vertices $u,v\in \overline{P}_{\ge k^*}$, define event
\begin{align*}
    \mathcal{A}_{uv}=\left\{\abs{\Gamma_1^{G_1}(u)\cap\Gamma_1^{G_2}(v)\cap \overline{P}_{\ge k^*}}\leq 2\right\},
\end{align*}
and $\calA=\bigcap_{u,v\in \overline{P}_{\ge k^*}:u\neq v}\calA_{uv}$.
By \prettyref{lmm:upperbound-P0-fake} and union bound, 
we have $\prob{\calA^c} \le n^{-1+o(1)}.$  

Second, let $\calB$ denote the event that all true pairs in $P_{k^*}$ are matched successfully. By \prettyref{lmm:proofPk}, $\prob{\calB}\ge 1-n^{-1.5+o(1)}$.

Third, by \prettyref{lmm:uinPk} and union bound, we have $\prob{Q_{\ge k^*}\subset \hat{Q}_{\ge k^*}\subset \overline{Q}_{\ge k^*}}\le n^{-4+o(1)}.$

Finally, by \prettyref{lmm:proofPGMtrue}, we have 
$$ 
\prob{\abs{S_{h^*}}\geq \frac{1}{2}n_{k^*+h^*}}\geq 1- n^{-3+o(1)}. 
$$
 Combining the above, it follows that 
$$
\prob{\calA \cap \calB\cap \{Q_{\ge k^*}\subset \hat{Q}_{\ge k^*}\subset \overline{Q}_{\ge k^*}\} \cap \{ \abs{S_{h^*}}\geq \frac{1}{2}n_{k^*+h^*} \} }\ge 1-n^{-1+o(1)}.
$$

Now, suppose event $\calA \cap \calB\cap \{Q_{\ge k^*}\subset \hat{Q}_{\ge k^*}\subset \overline{Q}_{\ge k^*}\}\cap \{ \abs{S_{h^*}}\geq \frac{1}{2}n_{k^*+h^*} \}$ holds. 
We aim to show that $\calR_{k^*+1}$ contains 
no fake pair in $\hat{Q}_{\ge k^*}$ and 
all true pairs $(u,u)$ with $u \in S_h$ for $h \ge 0.$


We first show $\calR_{k^*+1}$ contains 
no fake pair in $\hat{Q}_{\ge k^*}$. Suppose not. 
Let  $(u,v)$ denote the first fake pair in $\hat{Q}_{\ge k^*}$ matched by the PGM algorithm. This implies that the PGM only matches true pairs before matching $(u,v)$. 
Since the threshold $r$ of the PGM is set to be $3$, 
it follows that $(u,v)$ has at least three $1$-hop witnesses 
that are true pairs in $\hat{Q}_{\ge k^*}$.  
Since $\hat{Q}_{\ge k^*}\subset \overline{Q}_{\ge k^*}$,
it follows that 
$\abs{\Gamma_1^{G_1}(u)\cap\Gamma_1^{G_2}(v)\cap \overline{P}_{\ge k^*}}\geq 3$,
which contradicts the fact that event $\calA$ holds. 
Thus, $\calR_{k^*+1}$ contains no fake pairs in $\hat{Q}_{\ge k^*}$.

Next, we prove that $\calR_{k^*+1}$ contains
all true pairs in $S_h$ for all $h\ge 0$ by induction.
For ease of presentation, 
we assume $\calR_{k^*+1}$ contains the match pairs in the previous slice,
that is $\calR_{k^*+1} \supset \calR_{k^*}.$
The base case with $h=0$ follows from the definition of $\calB$. 
Assume the induction hypothesis holds for $h-1$, we aim to show it continues to hold for $h.$  Based on the definition of $S_h$, the true pairs in $S_h$ have at least 3 common 1-hop neighbors in $S_{h-1}$. Because all true pairs in $S_{h-1}$ have been matched and $Q_{\ge k^*}\subset\hat{Q}_{\ge k^*}$, the true pairs in $S_h$ would be matched by the PGM algorithm with threshold $r=3$. Therefore, $\calR_{k^*+1}$ contains
all true pairs in $S_h$  for all $h\ge 0$.

Finally,
$$
\abs{S_{h^*}}\ge \frac{1}{2}n_{k^*+h^*}=\frac{C}{2}n \alpha^{1-\beta}_{k^*+h^*}\ge \frac{C}{2}n (2\tilde{w})^{1-\beta},
$$
where 
$\tilde{w}=\left(\frac{192\overline{w}\ln 2}{Cs^2}\right)^{1/(3-\beta)}= \Theta(1)$
and the last inequality holds due to
the choice of $h^*$. 
Thus, $\calR_{k^*+1}$ has $\Theta(n)$ true pairs.

\subsubsection{Proof of \prettyref{lmm:lowerbound-P0-true}}\label{sec:pf-lmm9}
Fix a vertex $u\in P_0$. For each vertex $i\in P_{k^*}$,  let $x_i$ be a binary random variable such that $x_i=1$ if $i$  connects to $u$ both in $G_1$ and $G_2$, and $x_i=0$ otherwise.  Then,  $\abs{\Gamma_{1}^{G_1}(u)\cap \Gamma_{1}^{G_2}(u)\cap P_{k^*}}=\sum_{i \in P_{k^*}} x_i$
 and $x_i$'s are independent. Moreover, we have
 \begin{align*}
    \prob{x_i=1}\geq \frac{\alpha_{k^*}\alpha_{0}}{n\overline{w}}s^2.
\end{align*}
Recall $\abs{P_{k^*}}\geq n_{k^*}= C n \alpha_{k^*-1}^{1-\beta}$ in view
of \prettyref{eq:Pknumber}. Hence,
$$\abs{\Gamma_{1}^{G_1}(u)\cap \Gamma_{1}^{G_2}(u)\cap P_{k^*}}\overset{s.t.}{\geq}\Binom\left(n_{k^*},\frac{\alpha_{0}\alpha_{k^*}}{n\overline{w}}s^2\right).$$

Thus, we apply Chernoff Bound in \prettyref{thm:chernoffbound} with $\eta=\frac{1}{2}$ and get
\begin{align*}
   \prob{\abs{\Gamma_{1}^{G_1}(u)\cap \Gamma_{1}^{G_2}(u)\cap P_{k^*}}\leq \frac{C\alpha_{k^*}^{2-\beta}\alpha_0s^2}{8\overline{w}} }\le\prob{\Binom\left(n_{k^*},\frac{\alpha_{0}\alpha_{k^*}}{n\overline{w}}s^2\right)\leq \frac{C\alpha_{k^*}^{2-\beta}\alpha_0s^2}{2^{\beta}\overline{w}} }\le n^{-4},
\end{align*} 
where the last inequality holds because $n_{k^*}\frac{\alpha_{k^*}\alpha_{0}}{n\overline{w}}s^2 =\frac{C\alpha_{k^*}^{2-\beta}\alpha_0s^2}{2^{\beta-1}\overline{w}} \geq 64\log n$,
due to the choice of $k^*$ in~\prettyref{eq:choice_k_star}.

\subsubsection{Proof of \prettyref{lmm:upperbound-P0-fake}}\label{sec:pf-lmm10}
Fix two distinct vertices $u,v\in \overline{P}_0$. We bound from above
the number of their common $1$-hop neighbors in $\overline{R}=\bigcup_{k\ge 1}\overline{P}_k$.

For each $k \ge 1$ and each vertex $i\in \overline{P}_{k}$,  let $y_i^k$ be a binary random variable such that $y_i^k=1$  if $i$ is connected to $u$ in $G_1$ and $v$ in $G_2$, and $y_i^k=0$ otherwise. 
Since the event that $i$ is connected to $u$ is independent of the event that $i$ is connected to $v$, we have 
\begin{align*}
    \prob{y_i^k=1}\leq \left(\frac{(1+2\delta)\alpha_{k-1}w_{\max}}{n\overline{w}}s\right)^2\leq (1+2\delta)^2\frac{\alpha_{k-1}^2}{n\overline{w}}s^2 \triangleq p_{\max}^k, \quad \forall k \ge 1.
\end{align*}

Moreover, $y_i^k$'s are independent. Thus,
$$\abs{\Gamma_{1}^{G_1}(u)\cap \Gamma_{1}^{G_2}(v)\cap \overline{R}}\overset{s.t.}{\leq} \sum_{k=1}^{K}\Binom \left(\abs{\overline{P}_k},p_{\max}^k\right).$$

Recall  $n_k=Cn\alpha^{1-\beta}_{k-1}$ in view of \prettyref{eq:Pknumber}, $n_k\le\abs{\overline{P}_k}\leq 2n_k$,
and $\kappa =\frac{(1+2\delta)^22^{5-\beta}C}{(2^{3-\beta}-1)\overline{w}}$.
Thus, 
\begin{align*}
&\sum_{k=1}^{K}\abs{\overline{P}_k}p_{\max}^k\le\sum_{k=1}^{K}2n_{k}\frac{(1+2\delta)^2\alpha_{k-1}^2}{n\overline{w}}s^2
= \frac{2Cn^{\gamma(3-\beta)}s^2}{\overline{w}}\sum_{k=1}^{K}\frac{(1+2\delta)^2}{2^{(k-1)(3-\beta)}} \leq 2\kappa n^{\gamma(3-\beta)}s^2,\\
&\sum_{k=1}^{K}\abs{\overline{P}_k}p_{\max}^k\ge n_1\frac{\alpha_{0}^2}{n\overline{w}}s^2= \frac{Cn^{\gamma(3-\beta)}}{\overline{w}}s^2\ge 64\log n.
\end{align*}
Then, we apply Chernoff Bound  in \prettyref{thm:chernoffbound} with $\eta=1$, and get
\begin{align*}
    & \prob{\abs{\Gamma_{1}^{G_1}(u)\cap \Gamma_{1}^{G_2}(v)\cap \overline{R}}\geq 4\kappa n^{\gamma(3-\beta)}s^2}\le\prob{\sum_{k=1}^{K}\Binom \left(\abs{\overline{P}_k},p_{\max}^k\right)\geq 4\kappa n^{\gamma(3-\beta)}s^2} \le n^{-4}.
\end{align*}

\subsubsection{Proof of \prettyref{lmm:proofP0}}\label{sec:pf-lmm11}
Recall the bound of the number of 1-hop witnesses is provided by \prettyref{lmm:lowerbound-P0-true} and \prettyref{lmm:upperbound-P0-fake}.

First, for any vertex $u\in P_0$, define event
\begin{align*}
    \mathcal{A}_u=\left\{\abs{\Gamma_1^{G_1}(u)\cap\Gamma_1^{G_2}(u)\cap P_{k^*}}\geq \frac{C\alpha_{k^*}^{2-\beta}\alpha_0s^2}{2\overline{w}}\right\},
\end{align*}
and $\calA=\bigcap_{u\in P_0}\calA_u$. By \prettyref{lmm:lowerbound-P0-true} and union bound, we have $
  \prob{\mathcal{A}}\leq n^{-3+o(1)}.$

Second, for any two distinct vertices $u,v\in \overline{P}_0$, define event
\begin{align*}
    \mathcal{B}_{uv}=\left\{\abs{\Gamma_1^{G_1}(u)\cap\Gamma_1^{G_2}(v)\cap \overline{R}}\leq 4\kappa n^{\gamma(3-\beta)}s^2\right\},
\end{align*}
and $\calB=\bigcap_{u,v\in \overline{P}_0:u\neq v}\calB_{uv}$.
By \prettyref{lmm:upperbound-P0-fake} and union bound, 
we have $\prob{\calB^c} \le n^{-2+o(1)}.$

Third, we define an event 
$
\calC =\bigcap_{0 \le k \le k^*} \left\{Q_k\subset \hat{Q}_k \subset \overline{Q}_k\right\}\cap \left\{Q_{\ge k^*}\subset \hat{Q}_{\ge k^*} \subset \overline{Q}_{\ge k^*}\right\}.
$
By \prettyref{lmm:uinPk} and union bound, we have $\prob{\calC^c}\le n^{-4+o(1)}.$

Finally, we let $\calE$ denote the event that $\hat{\calR}$ contains all true pairs in $Q_{k^*}$ and no fake pairs in $\hat{Q}_{k}$ for any $k\ge 1$. 
By~\prettyref{lmm:proofP1}, \prettyref{lmm:proofPk} and \prettyref{lmm:proofPGM}, $\prob{\calE^c} \le n^{-1.5+o(1)}.$

Combining the above, it follows that 
$$
\prob{ \calA \cap \calB \cap \calC \cap \calE } \ge 1- n^{-3+o(1)}-n^{-2+o(1)}  - n^{-4+o(1)} -n^{-1.5+o(1)}
\ge 1-n^{-1.5+o(1)}.
$$
Suppose $\calA \cap \calB \cap \calC \cap \calE$ holds.
Then, $\hat{\calR}$ contains all true pairs in $Q_{k^*}$, and thus the minimum number of $1$-hop witnesses among all true pairs $(u,u)$ in $Q_0\subset \hat{Q}_0$ is lower bounded by $\frac{C\alpha_{k^*}^{2-\beta}\alpha_0s^2}{8n\overline{w}}.$
Moreover, since $\hat{\calR}$ contains no fake pairs in $\hat{Q}_{\ge 1}$ and 
$\hat{Q}_{\ge 1} \subset \overline{Q}_{\ge 1}$ on event $\calC$, 
it follows that $\hat{\calR}$ is contained by all the true pairs in $\bigcup_{k\ge1}\overline{Q}_k$, \ie, all the true pairs with weights no larger than $(1+2\delta)n^{\gamma}$.
Thus, 
the maximum number of $1$-hop witnesses among all fake pairs $(u,v)$ in $\hat{Q}_0\subset \overline{Q}_0$
is upper bounded by 
$4\kappa n^{\gamma(3-\beta)}s^2.
$
Note that by the choice of $k^*$ in~\prettyref{eq:choice_k_star}, $\frac{C\alpha_{k^*}^{2-\beta}\alpha_0s^2}{8\overline{w}}\ge \frac{Cn^{\gamma}s^2}{8\overline{w}}\left(\frac{192\overline{w}\log n}{Cs^2}\right)^{\frac{2-\beta}{3-\beta}} >4\kappa n^{\gamma(3-\beta)}s^2$,
where the last inequality hols for all sufficiently large $n$ in view of $2 <\beta<3$. 
Moreover,
since $\overline{P_0} \subset P_0 \cup P_1$,
 for any fake pair $(u,v) \in \hat{Q}_0$, the two corresponding true pairs $(u,u),(v,v) \in Q_0 \cup Q_1.$
 Therefore, the two true pairs either have more $1$-hop witnesses than the
 fake pair $(u,v)$ or have already  been matched in $\hat{Q}_1.$
 Hence, $\calR_0$ contains 
all true pairs in $Q_0$ and
no fake pairs in $\hat{Q}_0$.

\bibliography{matching}
\end{document}